\documentclass[prd,aps,tightenlines,superscriptaddress,nofootinbib,preprint]{revtex4}
\usepackage[colorlinks=true, pdfstartview=FitV, linkcolor=red, citecolor=magenta, urlcolor=magenta]{hyperref}
\usepackage{amsmath,amsfonts,amsthm,amssymb}
\usepackage{caption}
\usepackage{mathtools}
\usepackage{multirow}
\usepackage{graphicx}
\usepackage{xcolor}
\usepackage{color}
\usepackage{epsfig}
\usepackage{amsmath}
\usepackage{comment}
\usepackage{nameref}
\usepackage{cleveref}
\usepackage{caption}
\usepackage{subcaption}
\usepackage{orcidlink}
\usepackage{verbatim}
\usepackage{comment}
\theoremstyle{definition}

\begin{document}
\title{Thermodynamic and observational implications of black holes in toroidal geometry}

\author{Usman Zafar \orcidlink{0000-0001-9610-1081} \footnote{\color{blue}zafarusman494@gmail.com,~s2471001@ipc.fukushima-u.ac.jp}}\affiliation{Faculty of Symbiotic Systems Science, Fukushima University, Fukushima 960-1296, Japan}
\author{Kazuharu Bamba  \orcidlink{0000-0001-9720-8817}\footnote{\color{blue}bamba@sss.fukushima-u.ac.jp}}
\affiliation{Faculty of Symbiotic Systems Science, Fukushima University, Fukushima 960-1296, Japan}
\author{Abdul Jawad \orcidlink{0000-0001-5249-803X}\footnote{\color{blue}abduljawad@cuilahore.edu.pk}}
\affiliation{Department of Mathematics, COMSATS University
Islamabad, Lahore-Campus, Lahore-54000, Pakistan}
\author{Tabinda Rasheed  \orcidlink{0009-0001-5513-1513}\footnote{\color{blue}tabindarasheed00@gmail.com,~202451150001@nuist.edu.cn}}
\affiliation{School of Mathematics and Statistics, Nanjing University of Information Science and Technology, Nanjing 210044, China}
\author{Sanjar Shaymatov \orcidlink{0000-0002-5229-7657}\footnote{\color{blue}sanjar@astrin.uz}}
\affiliation{Power Engineering Faculty, Tashkent State Technical University, Tashkent 100095, Uzbekistan}\affiliation{Faculty of Information Technologies, University of Tashkent for Applied Sciences, Str. Gavhar 1, Tashkent 100149, Uzbekistan}
 \affiliation{School of High Technologies and Innovative Engineering, Western Caspian University, Baku AZ1001, Azerbaijan}

\begin{abstract}
We investigate the thermodynamic and observational implications for the charged torus-like black holes, a class of solutions distinct from the classical Schwarzschild black holes.  We explicitly derive the fundamental thermodynamic properties, such as heat capacity, P-V diagram, isothermal compressibility, Helmholtz free energy, and Gibbs free energy, under different entropy models. We find that only the exponential corrected entropy demonstrates multiple phase transitions, which we validate with the Ricci Scalar divergence obtained from the Ruppeiner formalism. This indicates that exponential corrected entropy is more sensitive to BH's microstructure as compared to the Hawking-Bekenstein and R\'{e}nyi entropy models within the non-spherical (toroidal) horizon. In addition, we study the sparsity and emission rates of Hawking radiation, demonstrating that exponential correction entropy yields more consistent and stable behavior. In our observational analysis, we graphically demonstrate the behavior of redshift, blueshift, and gravitational shift, and identify specific conditions where the photon sphere radius exceeds the innermost stable circular orbit radius, which depends on the values of parameters such as electric charge and cosmological constant. The novel insight of this work is that despite this violation, our computed redshift, blueshift, and gravitational shifts fall within the range of the observational data of NGC 4258 and UGC 3789. 
\end{abstract}

\maketitle

\section{Introduction}
Black holes (BHs) have become one of the most important topics in the theory of general relativity, which provides a very unique framework where the concepts of thermodynamics, gravity, and observational astrophysics align. Hawking and Bekenstein's foundational work on BH thermodynamics has established a significant link between gravitational theory and quantum mechanics \cite{Hawking:1974rv,Bekenstein:1973ur,Bekenstein:1974ax,Hawking:1975vcx,Bekenstein:1972tm,Bardeen:1973gs}. In this framework, BHs follows the laws of thermodynamics, where surface gravity corresponds to temperature (Hawking temperature), the area of the event horizon signifies entropy and exhibiting phase transition akin to those in conventional thermodynamic system \cite{Bardeen:1973gs}. Analytical frameworks like $P-V$ criticality \cite{Chamblin:1999tk,Chamblin:1999hg,Kubiznak:2012wp,Kubiznak:2016qmn,Karch:2015rpa,Mancilla:2024spp}, stability analysis utilizing  heat capacity, Gibbs free energy (GFE) and Helmholtz free energy (HFE) \cite{Rodrigues:2022qdp,Simovic:2023yuv,Ali:2019myr,Rani:2025esb}, provide significant insights on BH characteristics and their thermodynamic behavior.  For example, a BH is considered locally and globally stable if the heat capacity, HFE, and GFE demonstrate positive behavior and vice versa (for further details regarding local and global stability of BHs, check Refs.~\cite{Rodrigues:2022zph,Dehyadegari:2017flm, Wei:2020poh,Liu:2014gvf,Dehyadegari:2020tau}). 

In BH thermodynamics, different types of phase transition have been explore such as Davies-type phase transition occurs due to the heat capacity divergence \cite{Davies:1989ey,Lousto:1994jd,Muniain:1995ih}, the Hawking-Page phase transition between thermal Anti-de-Sitter (AdS) spacetime and BHs \cite{Hawking:1982dh}, extremal phase transition \cite{Pavon:1988in,Pavon:1991kh,Cai:1996df,Cai:1998ep,Wei:2009zzf,Bhattacharya:2019awq}, and Van-dar-Waal  like behavior in the extended phase space, where cosmological constant serves as thermodyanmic pressure and mass interpreted as enthalpy \cite{Kastor:2009wy,Dolan:2010ha,Dolan:2011xt,Dolan:2011jm,Dolan:2012jh,Kubiznak:2016qmn,Bhattacharya:2017nru}. 

In this context, thermodynamic geometry provides an effective framework to study the BH phase transition by introducing various thermodynamic metrics, and these metrics are constructed in terms of entropy, which gives us valuable insight regarding the phase transition when the divergence of Ricci curvature scalar determined from this metric formalism aligns with the zero point (ZP) or divergence of heat capacity. The metric formalism introduced by the Weinhold \cite{Weinhold:1975xej}, based on equilibrium state space, which was later extended by the Ruppiener \cite{Ruppeiner:2008kd,Ruppeiner:1995zz}, whose formalism is closely linked with the Weinhold's formalism (for other thermodynamic geometry formalism and their constructions check Refs.~\cite{NaveenaKumara:2020biu,Quevedo:2008ry,Akbar:2011qw,Hendi:2015cka,Sarkar:2006tg,Bhattacharya:2019qxe,Hendi:2015xya,Banerjee:2016nse}), it reveals valuable insights regarding the microscopic interaction and phase transitions. Although research on non-spherical BH configurations has increased, toroidal or torus-like BHs have not been investigated as extensively as spherically symmetric ones. The main reason is that they have very intricate topology and horizon geometry, which makes them difficult to analyze.  These geometries naturally arise within AdS spacetime and offer a framework for studying the thermodynamic and observational characteristics. Departing from previous analyses, which focus on the hyperbolic and planar horizons, our toroidal framework features a novel dual periodic topology, providing modified thermodynamic curvature characteristics. In torus-like spacetimes, entropy correction plays a significant role because the traditional entropy  (Hawking-Bekenstein entropy) formulated under classical assumptions may not fully incorporate quantum or topological complexities. To better describe these systems, R\'{e}nyi \cite{Odintsov:2023qfj} and exponential corrected entropy \cite{ghosht}  introduce generic statistical or quantum modifications that may be fundamentally significant. 

Furthermore, at a constant temperature, Hawking radiation produces sparse and discrete quanta which deviate from the continuous emission pattern of classical blackbody radiation \cite{Page:2004xp,Gray:2015pma}. The nature of Hawking radiation significantly impacts both its detectability and quantum information aspects. Moreover, we also address the Hawking radiation by exploring emission rates and sparsity, which elucidate the effects of entropy corrections on BH evaporation.  In recent years, strong evidence supporting the existence of BHs has emerged from the recent observational phenomena, including the stars' orbital motion near the Galactic center \cite{Ghez:2008ms,Morris:2012uz,Eckart:1996zz,Gillessen:2008qv}, imaging of supermassive BHs in M87 and the Milky Way \cite{EventHorizonTelescope:2019ths,EventHorizonTelescope:2022xqj}, and the gravitational wave observation by LIGO-Virgo \cite{LIGOScientific:2016aoc,LIGOScientific:2016sjg}. These findings have encouraged the formulation of methods to determine the BH parameter by using observational data, for example, to determine properties like mass and spin, one widely used approach is to study the frequency shifts (red shift and blueshift) of photons emitted by orbiting particles, originally applied in Kerr geometries \cite{Herrera-Aguilar:2015kea,Banerjee:2022him,Momennia:2023lau}. Subsequent development has extended this methodology to a wider class of BH solutions, including Kerr-Newmann dS \cite{Kraniotis:2019ked}, higher dimensional \cite{Sharif:2016rhs}, regular \cite{Becerril:2016qxf}, polymerized types \cite{Fu:2022afk}, and has also been employed to compute the Hubble constant \cite{Momennia:2023lau}. Numerous works have been done to apply general relativity to real astrophysical systems, for example, active galactic nuclei (AGN) with water maser disks, for calculating the ratio between the BH mass and distance \cite{Nucamendi:2020tov,Villalobos-Ramirez:2022xic,Villaraos:2022euo,Gonzalez-Juarez:2022pya}. Recently, the concept of redshift rapidity, have made it possible to distinguish distance from mass in Reisner-nordst\"{o}rm (RN) and Schwarzschild spacetimes with the help of observational data \cite{Momennia:2023svi,Morales-Herrera:2024zbr}. The values of parameters such as electric charge and spin still exhibit ambiguity, specifically in different models of the modified theory of gravity \cite{Ghez:2008ms,Morris:2012uz,Eckart:1996zz,Gillessen:2008qv,EventHorizonTelescope:2019ths,EventHorizonTelescope:2022xqj,LIGOScientific:2016aoc,LIGOScientific:2016sjg}. When relativistic formalisms are extended to a general spherically symmetric spacetime, it allows us to investigate the impacts of additional parameters predicted by modified theories of gravity and analyze their deviation from Einstein's theory \cite{Herrera-Aguilar:2015kea,Banerjee:2022him,Momennia:2023lau}. We extended this general spherical symmetric formalism to discuss the frequency shifts (redshift, blueshift, and gravitational shift) for the torus-like BH solution, which has toroidal geometry. In addition, studying frequency shifts, which is commonly done for spherical BHs, is equally crucial for torus-like geometries, as it helps in examining strong-field lensing, accretion disk behavior, and deviations in trajectories of photons and orbital stability. The resulting deviations may lead to observable astrophysical signals, establishing frequency shifts, as an effective method for identifying non-traditional horizon geometries. We also demonstrate the observational significance of our framework for these types of BHs by comparing its predicted frequency shifts with measurements from megamaser systems. 

This paper is arranged as follows: in
\texttt{Sec.~\ref{MR}}, we discuss the torus-like BH solution in detail. We analytically probe the thermodynamic aspects of charged torus-like BHs in
\texttt{Sec.~\ref{TDB}}, where we derive thermodynamic quantities such as mass, temperature, and volume, analyze the heat capacity, $P-V$ diagram,  isothermal compressibility, GFE, and HFE in terms of various entropy models including Hawking-Bekenstein, R\'{e}nyi, and exponential corrected entropy frameworks. In
\texttt{Sec.~\ref{TDG}}, we utilize the thermodynamic geometry formalism to investigate the behavior of Ricci curvature scalar for the Hawking-Bekenstein, exponential corrected, and R\'{e}nyi entropies to identify the possible phase transition.
\texttt{Sec.~\ref{SHR}} provides a detailed analysis of the sparsity of the Hawking radiation and energy emission from charged torus-like BHs using the Hawking-Bekenstein, exponential corrected, and R\'{e}nyi entropies. Extending our thermodynamic analysis of charged torus-like BHs to the observational analysis, we analytically derive the behavior of a particle in time-like and null geodesics and employ these results to evaluate and graphically represent the frequency shift, including redshift, blueshift, and gravitational shift, in 
\texttt{Sec.~\ref{FRS}}. The conclusion of our work is presented in  \texttt{Sec.~\ref{CNC}}.                

\section{Charged Torus-Like BH: Metric Review}\label{MR}
In this section, we describe the metric solution for the charged torus-like BH. Toroidal BHs offer a topologically unique solution characterized by complex geometrical and thermodynamic features as compared to the spherically symmetric BHs, which are broadly studied as described in Refs.~\cite{Huang,Han,Ali:2023pyv,Feng:2021vey,Liang:2021elg}. These spacetimes are significant in the AdS framework, where the negative cosmological constant permits the horizon topology to differ from spherically symmetric spacetimes. In Ref.~\cite{Huang}, a static charged BH solution has been proposed, which is determined by solving the Einstein-Maxwell equation in the presence of a cosmological constant. Therefore, the action for this BH solution is given as 
\begin{eqnarray}\label{Action}
\mathcal{S}=\int \sqrt{-g}\left(-2\Lambda-F^{uv}F_{uv}+R\right)d^{4}x\,,
\end{eqnarray}
where $\sqrt{-g}$ is the determinant of the metric tensor $g_{uv}$, which ensures the general covariance of the action, and $R$ presents the curvature of spacetime obtained from the Riemann curvature tensor. Furthermore, $\Lambda$ presents the cosmological constant, which indicates dS for $\Lambda>0$ and AdS for $\Lambda<0$. The second term in Eq.~\eqref{Action} is $F^{uv}F_{uv}$, which presents the electromagnetic field computed from the electromagnetic four potential $\mathbb{A}_{u}$. As provided in Ref.~\cite{Huang}, the line element for a generic static torus-like BH can be written as 
\begin{eqnarray}\label{LE}
ds^{2}=-C(r)dt^{2}+D(r)dr^{2}+r^{2}(d\theta^{2}+d\phi^{2})\,,
\end{eqnarray}
where $C(r),~D(r)>0$ and ranges for angular coordinates are $\theta\in[0,~2\pi]$ and $\phi\in[0,~2\pi]$. If we look at Eq.~\eqref{LE}, it can be noticed that it possesses three space-like killing vectors in addition to $\partial/\partial{t}$, including $\partial/\partial{\theta},~\partial/\partial{\phi}$ and $\theta(\partial/\partial{\phi})-\phi(\partial/\partial{\theta})$ which indicates that this equation correspond to the general static spherically symmetric solution presented in \eqref{Action} as described in Refs.~\cite{Huang, Han}. Thereby, an explicit solution for the metric is obtained by using the field equation of Eq.~\eqref{Action}, which yields 
\begin{eqnarray}\label{1b}
ds^{2}=-f(r)dt^{2}+\frac{1}{f(r)}dr^{2}+r^{2}(d\theta^{2}+d\phi^{2})\,,
\end{eqnarray}
where $f(r)$ is a metric function which is given as
\begin{eqnarray}\label{2b}
f(r)=-\frac{2 M}{\pi  r}+\frac{4 Q^2}{\pi  r^2}-\frac{\Lambda
r^2}{3}\,.
\end{eqnarray}
Here, we represent $\Lambda$ as the cosmological constant where
$\Lambda =-8\pi  P$ (in extended thermodynamics, $\Lambda$ can be interpreted as the thermodynamic pressure; for more details, see Ref.~\cite{Kastor:2009wy}), $M$ is the mass of the BH, and $Q$ is the
charge of the torus-like BH. It is also important to note that this solution reduces to Kar's solution when the cosmological constant is set to zero \cite{Huang}. Moreover, BH's basic geometric as well as thermodynamic features are represented by the metric function $f(r)$, for example, if we put $r=0$ then $f(r)\to \infty$ indicating a curvature singularity. Similarly, for higher (or larger) values of $r$, the spacetime tends towards an asymptotically AdS framework owing to the term $-\Lambda r^{2}/3$. By plugging the value of the cosmological constant $\Lambda=-8\pi  P$ into Eq.~\eqref{2b}, it yields
\begin{eqnarray}\label{3b}
f(r)=-\frac{2 M}{\pi  r}+\frac{8}{3} \pi  P r^{2}+\frac{4 Q^2}{\pi  r^{2}}\,.
\end{eqnarray}

Now, we determine the mass for the torus-like BH by using Eq.~\eqref{3b} in the following relation $f(r_\mathrm{e})=0$, given as
\begin{eqnarray}\label{MBH}
M=\frac{2 \left(2 \ \pi ^2 P \ r^{4}_\mathrm{e}+3 \ Q^2\right)}{3 \ r_\mathrm{e}}\,,
\end{eqnarray}
where $r_\mathrm{e}$ corresponds to the radius of the event horizon of the charged torus-like BH.
\section{Thermodynamics Quantities of Torus-like BH Through the Corrected Entropies}\label{TDB}
In this section, we examine the impact of entropies that incorporate the logarithmic corrections, such as exponential and Rényi entropy, to analyze the thermodynamic quantities of the Torus-like BH. Before delving into our analysis of thermodynamic quantities, we first discuss the entropy frameworks that we utilize in this study. The Hawking-Bekenstein formula establishes the fundamental link between BH entropy and the surface area of its event horizon. Mathematically, it is expressed as 
\begin{eqnarray}\label{HB}
S_\mathrm{HB}=\frac{\mathcal{A}}{4 G}\,,
\end{eqnarray}
where $\mathcal{A}=4\pi^{2} r^{2}_\mathrm{e}$ presents the area of the horizon \cite{Feng:2021vey}, which is different from the spherical case, and $G$ is the gravitational constant. We mention here that this distinction is quite crucial for maintaining consistent thermodynamic parameters, such as entropy and volume, since the geometric factor $4\pi^{2}$ corresponds to the $S^{\star}\times S^{\star}$ horizon topology \cite{Ali:2023pyv} that we employ consistently in our thermodynamic analysis. The standard semiclassical expression for BH entropy is believed to be altered by quantum gravity, notably through exponential corrections that arise naturally in frameworks like non-local gravity, string theory, and quantum statistical approaches. Such exponential terms capture non-perturbative contributions and reveal crucial details about the microscopic framework of the spacetime. Furthermore, these terms may enhance the study of thermodynamic stability, phase transitions, and singularity resolution, effectively associating classical BH thermodynamics with foundational features of quantum gravity. In this perspective, by following the same methodology as described in Ref.~\cite{ghosht}, the expression for the exponential corrected entropy can be written as
\begin{eqnarray}\label{1ee}
 S_\mathrm{Exp}=\frac{\mathcal{A}\ln{2}}{8\nabla\pi{l^{2}_{p}}}+\exp\left(-\frac{\mathcal{A}\ln{2}}{8\nabla\pi{l^{2}_{p}}}\right)\,,
\end{eqnarray}
where $l^{2}_{p}=~1,~\mathcal{A}=4\pi^{2}{r^{2}_\mathrm{e}}$ and
$\nabla=\frac{\ln{2}}{2\pi}$. By inserting these values, one can easily derive the following form of exponential corrected entropy
\begin{eqnarray}\label{2ee}
S_\mathrm{Exp}=\exp\left(-\pi^{2}r^{2}_\mathrm{e}\right)+\pi^{2}r^{2}_\mathrm{e}\,.
\end{eqnarray}

Constantino Tsallis developed non-extensive entropy as a generalization of Boltzmann-Gibbs entropy, making it suitable for analyzing systems that exhibit non-linear behavior and strong sensitivity to initial conditions. Whereas Boltzmann-Gibbs entropy is based on the proportional scaling of entropy with system size, non-extensive entropy addresses situations in which this scaling principle is violated. For example, the gravitational systems like stars, BHs, and galaxies, where the long-range nature of gravity leads to correlations between particles over large distances, violate the additive property assumed in Boltzmann-Gibbs entropy. The concept of non-extensive entropy is extensively utilized in statistical mechanics, cosmology, and theoretical physics, notably in contexts involving fractal structures, memory phenomena, or long-range forces \cite{Odintsov:2023qfj}. Mathematically, the R\'{e}nyi entropy can be expressed as 
\begin{eqnarray}\label{RE}
S_\mathrm{R}=\frac{1}{\alpha}\ln{\left(1+\alpha \ S_\mathrm{HB}\right)}\,,
\end{eqnarray}
where $\alpha$ depicts the non-extensive parameter. The range $0<\alpha<1$, ensures that the R\'{e}nyi entropy $S_\mathrm{R}$ is mathematically consistent and thermodynamically viable, whereas values beyond this limit lead to an undefined form because of the entropy’s convex behavior (for more details in terms of considering non-extensive parameter $\alpha=0$ and other non-extensive entropy framewoks, see Refs.~\cite{Odintsov:2023qfj,Nojiri:2022aof,Nojiri:2022dkr,Nojiri:2022sfd,Elizalde:2025iku,Odintsov:2023vpj,Nojiri:2024zdu}). Thereby, by comparing these above mention entropy models enables us to examine how quantum correction and non-extensive effects modifies the thermodynamic phase structure of BHs with complex topology.  

It is important to mention that the thermodynamics models motivated by the quantum gravity indicate that the generalized entropy given in Refs.~\cite{Nojiri:2022aof,Nojiri:2022dkr,Nojiri:2022sfd,Elizalde:2025iku,Odintsov:2023vpj,Nojiri:2024zdu,Odintsov:2022qnn} provides the most comprehensive parametrization of horizon entropy, incorporating the logarithmic, power-law, and non-perturbative deviations from the Bekenstein-Hawking area law. The generality of this framework ensures that R\'{e}nyi, Hawking-Bekenstein, Sharma-Mittal, Barrow, Tsallis, Loop quantum gravity, and Kindeski entropies arise with its specific limits. Despite offering significant mathematical stability, the generalized framework contains numerous free parameters that cannot be uniquely fixed, complicating the calculation of thermodynamic quantities and introducing ambiguity in thermodynamic studies. Therefore, using this generalized entropy framework makes our thermodynamic analysis more intricate, and it is also not analytically invertible. 

For these reasons, in this paper, we primarily focus on these representative and analytically tractable entropy frameworks, which are indeed the subclasses of the generalized entropy given in Refs.~\cite{Nojiri:2022aof,Nojiri:2022dkr,Nojiri:2022sfd,Elizalde:2025iku,Odintsov:2023vpj,Nojiri:2024zdu,Odintsov:2022qnn}; the Hawking-Bekenstein and R\'{e}nyi entropies, which capture the non-extensive effects associated with horizons with non-trivial topology, and the exponential corrected entropy, which accounts for quantum gravity corrections. Each of these entropy expressions arises from a specific limit of the generalized entropy framework, while still maintaining the analytic mapping $S({r_\mathrm{e}})\leftrightarrow r_\mathrm{e}(S)$ (which is quite intricate in the case of generalized entropy), allowing us to derive the expression for mass, conjugate temperature with respect to the corresponding entropy, heat capacity, P-V diagram, isotherm compressibility, free energies (HFE and GFE), and Ruppeiner curvature. Also, since in our analysis we want to validate the divergence in heat capacity using the thermodynamic geometry metric formalisms, the generalized entropy complicates the derivation and validation. Thus, the choice of entropy models in our analysis maintains physical comprehensiveness while remaining numerically tractable, enabling us to make a clear and straightforward comparison of thermodynamic features corresponding with the Hawking-Bekenstein, R\'{e}nyi, and exponential corrected entropies in the case of charged torus-like BHs.  
\subsection{Thermodynamic Mass}
Here, we determine the thermodynamic mass in terms of these entropy formalisms for the charged torus-like BH\footnote{We want to mention here that the thermodynamics of charged torus-like BHs in terms of Hawking-Bekenstein entropy have been discussed in Refs.~\cite{Ali:2023pyv,Feng:2021vey,Liang:2021elg,Ditta:2023gin}, but our aim is to discuss the impact of entropy corrections in comparison with the traditional entropy (Hawking-Bekenstein entropy). Therefore, we have analytically obtained the thermodynamic quantities in terms of the Hawking-Bekenstein, exponential correct, and R\'{e}nyi entropy models to enhance our understanding regarding the thermodynamic behavior of charged torus-like BHs.}. Firstly, we compute the mass of the BH in terms of the Hawking-Bekenstein entropy formalism by using Eqs.~\eqref{MBH} and \eqref{HB}, which yields 
\begin{eqnarray}\label{MHB}
M(S_{\mathrm{HB}},~P,~Q)=\frac{2 \pi} {3 \sqrt{S}} \left(\frac{2 P S^2}{\pi ^2}+3 Q^2\right)\,.
\end{eqnarray}
In the exponential entropy case, the BH’s mass is computed by using Eqs.~\eqref{MBH} and \eqref{2ee}. Consequently, the expressions of mass for the exponential corrected is given as
\begin{eqnarray}\label{Mee} M(S_{\mathrm{Exp}},~P,~Q)&=&\frac{2^{3/4} \pi} {3 \sqrt[4]{S-1}} \left(\frac{4 P (S-1)}{\pi ^2}+3 Q^2\right)\,.
\end{eqnarray}
Similarly, in the case of R\'{e}nyi entropy, one can easily derive the mass of the BH by employing Eqs.~\eqref{MBH} and \eqref{RE}. So, the expression of mass in terms of the  R\'{e}nyi entropy can be written as 
\begin{eqnarray}\label{MRE} 
M(S_{\mathrm{R}},~P,~Q)=\frac{2 \pi  \sqrt{\alpha }}{3 \sqrt{e^{\alpha  S}-1}} \left\{\frac{2 P \left(e^{\alpha  S}-1\right)^2}{\pi ^2 \alpha ^2}+3 Q^2\right\}\,.
\end{eqnarray}

Here, we emphasize that $M(S_{\mathrm{Exp}},~P,~Q)$ and $M(S_{\mathrm{R}},~P,~Q)$ are not computed by directly plugging $S_\mathrm{Exp}$ and $S_\mathrm{R}$ into Eq.~\eqref{MBH} (the Hawking-Bekenstein formula); instead, their derivation aligns with fundamental thermodynamic laws. We initiate the analysis of each entropy model by constructing their expressions as functions of the BH horizon radius $r_\mathrm{e}$; exponential corrected entropy is developed using the methodology provided in Ref.~\cite{ghosht}, and the R\'{e}nyi entropy is formulated via a non-extensive and non-additive approach from Ref.~\cite{Odintsov:2023qfj}. These expressions are subsequently rearranged to solve for $r_\mathrm{e}$ in terms of $S_\mathrm{Exp}$ or $S_\mathrm{R}$, respectively. By incorporating the inverted expressions introduced earlier into the torus-like BH’s metric function $f(r)$, the mass is re-expressed as a function of the corresponding entropy model. Thereby, $M_\mathrm{Exp}$ depends on the $S_\mathrm{Exp}$ and $M_\mathrm{R}$ depends on the $S_\mathrm{R}$. The approach complies with the first law of BH thermodynamics, thereby maintaining the physical validity of the derived results based on the chosen entropy model.
\subsection{Temperature}
This subsection presents the derivation of temperature, volume, and pressure, along with the formulation of the first law of thermodynamics for BHs under different entropy frameworks. It is straightforward to obtain the thermodynamic quantities conjugate to the Hawking-Bekenstein entropy by using Eq.~\eqref{MHB}, which is presented as
\begin{eqnarray}\nonumber
T_{\mathrm{HB}}=\frac{\partial M_\mathrm{HB}}{\partial S_\mathrm{HB}}&=&\frac{2 P \sqrt{S}}{\pi }-\frac{\pi  Q^2}{S^{3/2}}\,,\\\nonumber V_{\mathrm{HB}}=\frac{\partial M_\mathrm{HB}}{\partial P}&=&\frac{4 S^{3/2}}{3 \pi }\,,\\\label{THB} \phi_\mathrm{HB}=\frac{\partial M_\mathrm{HB}}{\partial Q}&=&\frac{4 \pi  Q}{\sqrt{S}}\,.
\end{eqnarray}

Before detailing the formulas for pressure and temperature within the exponential and R\'{e}nyi entropy approaches, the thermodynamic principles that underpin these derivations need to be properly explained. Under the generalized thermodynamic approach, different entropy formulations generate distinct thermodynamic phase spaces, compelling the formulation of BH mass within the context of each specific entropy model. As a result, the first law of BH thermodynamics in terms of exponential entropy can be altered accordingly, which is given as  
\begin{eqnarray}\label{FLBHE}
dM_\mathrm{Exp}=T_\mathrm{Exp}dS_\mathrm{Exp}+V_\mathrm{Exp}dP+\phi_\mathrm{Exp}dQ\,.
\end{eqnarray}
This formulation ensures that temperature $T_\mathrm{Exp}$ is always treated as the conjugate variable to entropy $S_\mathrm{Exp}$. Therefore, by using Eq.~\eqref{Mee}, we derive the conjugate temperature corresponding to the exponential entropy, which is given as 
\begin{eqnarray}\nonumber T_{\mathrm{Exp}}=\frac{\partial M_\mathrm{Exp}}{\partial S_\mathrm{Exp}}&=&\frac{4 P (S-1)-\pi ^2 Q^2}{2 \sqrt[4]{2} \pi  (S-1)^{5/4}}\,,\\\nonumber V_\mathrm{Exp}=\frac{\partial M_\mathrm{Exp}}{\partial P}&=&\frac{4\ \times 2^{3/4} (S-1)^{3/4}}{3 \pi }\,, \\\label{Tee}
\phi_\mathrm{Exp}=\frac{\partial M_\mathrm{Exp}}{\partial Q}&=&\frac{2\ 2^{3/4} \pi  Q}{\sqrt[4]{S-1}}\,.
\end{eqnarray}

Similarly, the first law of BH thermodynamics is modified in terms of R\'{e}nyi entropy, which is given as 
\begin{eqnarray}
dM_\mathrm{R}=T_\mathrm{R}dS_\mathrm{R}+V_\mathrm{R}dP+\phi_\mathrm{R}dQ\,.\label{FLBHER}
\end{eqnarray}
We compute the thermodynamic temperature, volume, and potential associated with the R\'{e}nyi entropy for the charged torus-like BH, which are expressed as
\begin{eqnarray}\nonumber
T_{\mathrm{R}}=\frac{\partial M_\mathrm{R}}{\partial S_\mathrm{R}}&=&\frac{e^{\alpha  S} \left(2 P \left(e^{\alpha  S}-1\right)^2-\pi ^2 \alpha ^2 Q^2\right)}{\pi  \sqrt{\alpha } \left(e^{\alpha  S}-1\right)^{3/2}}\,,\\\nonumber V_{\mathrm{R}}=\frac{\partial M_\mathrm{R}}{\partial V_\mathrm{R}}&=&\frac{4 \left(e^{\alpha  S}-1\right)^{3/2}}{3 \pi  \alpha ^{3/2}}\,,\\\label{TRE} \phi_\mathrm{R}=\frac{\partial M_{R}}{\partial Q}&=&\frac{4 \pi  \sqrt{\alpha } Q}{\sqrt{e^{\alpha  S}-1}}\,.
\end{eqnarray}
The foundation of the above-mentioned formalism lies in modern gravitational thermodynamics, which employs generalized entropic schemes such as the exponential corrected entropy and R\'{e}nyi entropy, capturing non-extensive and quantum gravitational features, respectively \cite{Rani:2025esb}. We mention here that to determine $T_\mathrm{Exp}$ and $T_\mathrm{R}$, we employ a robust method involving the differentiation of the BH mass with respect to the relevant entropy functions. This ensures that our thermodynamic model is physically consistent and aligns with the foundational principles of generalized statistical mechanics.

Importantly, employing a generalized entropy does not imply any correction or modification to the Hawking temperature, which stays determined entirely by the surface gravity through the relation $T_\mathrm{H}=\kappa/2\pi$. As emphasized in  Ref.~\cite{Nojiri:2021czz}, the inconsistency originates from substituting a non-Hawking-Bekenstein entropy into the first law without modifying the usual Bekenstein-Hawking mass-radius relation. By applying this procedure, the temperature becomes inaccurate since the function of mass is no longer conjugate to the modified entropy model within thermodynamics. In our approach, we avoid this inconsistency by computing the thermodynamic mass for each entropy model directly from the horizon condition and subsequently evaluating the temperature by using the first law formula given in Eqs.~\eqref{FLBHE} and \eqref{FLBHER}. Regardless of whether the entropy is the Hawking-Bekenstein, R\'{e}nyi, or exponential corrected, we first reconstruct the $M(S)$ by utilizing the specific entropy-radius relation and then determine the conjugate temperature associated with the corresponding entropy by applying the relation given in Eqs.~\eqref{THB},~\eqref{Tee}, and \eqref{TRE}. This guarantees that the temperature used in our thermodynamic treatment is consistent with the generalized entropy, resolving the inconsistency highlighted in Ref.~\cite{Nojiri:2021czz}, and this approach is adopted for different BHs as given in Refs .~\cite {Zafar:2025sxl,Nakarachinda:2022gsb,Bhattacharjee:2025oje,Yasir:2024zgm,Yasir:2025uov,Rani:2025esb,Tariq:2025wiy,Toledo:2019amt,Bhattacharya:2024bjp,Ladghami:2024sen}.

\subsection{Local Stability}
In this subsection, we will discuss the local stability of the charged torus-like BH, which is associated with the heat capacity. In the context of BH thermodynamics, heat capacity is a fundamental quantity that reflects stability if it is positive and signals instability when it is negative. Therefore, by following the procedure given in Refs.~\cite{Rodrigues:2022qdp,Davies:1989ey}, one can obtain the heat capacity relation for Hawking-Bekenstein entropy
\begin{eqnarray}\label{CHB}
C_{P}(S_{\mathrm{HB}})=T_\mathrm{HB}\left(\frac{\partial S_\mathrm{HB}}{\partial T_\mathrm{HB}}\right)\Bigg|_{P,~Q}&=&S \left(2-\frac{8 \pi ^2 Q^2}{2 P S^2+3 \pi ^2 Q^2}\right)\,.
\end{eqnarray}
One can gain a valuable understanding of BH stability and their thermodynamic responses to small perturbations by investigating their heat capacity. By using Eq.~\eqref{Tee}, the heat capacity $C_{P}$ corresponding to the exponential corrected entropy is derive as follows
\begin{eqnarray}\label{Cee} C_{P}(S_{\mathrm{Exp}})=T_\mathrm{Exp}\left(\frac{\partial S_\mathrm{Exp}}{\partial T_\mathrm{Exp}}\right)\Bigg|_{P,~Q}&=&\frac{4 (S-1) \left(\pi ^2 Q^2-4 P (S-1)\right)}{4 P (S-1)-5 \pi ^2 Q^2}\,.
\end{eqnarray}
Similarly, we compute the heat capacity associated with the R\'{e}nyi entropy by utilizing Eq.~\eqref{TRE} as
\begin{eqnarray}\label{CRE} 
C_{P}(S_{\mathrm{R}})=T_\mathrm{R}\left(\frac{\partial S_\mathrm{R}}{\partial T_\mathrm{R}}\right)\Bigg|_{P,~Q}&=&\frac{2 \left(e^{\alpha  S}-1\right) \left(2 P \left(e^{\alpha  S}-1\right)^2-\pi ^2 \alpha ^2 Q^2\right)}{2 \alpha  P \left(e^{\alpha  S}-1\right)^2 \left(3 e^{\alpha  S}-2\right)+\pi ^2 \alpha ^3 Q^2 \left(e^{\alpha  S}+2\right)}\,.
\end{eqnarray}

\begin{figure}[t] \centering
\epsfig{file=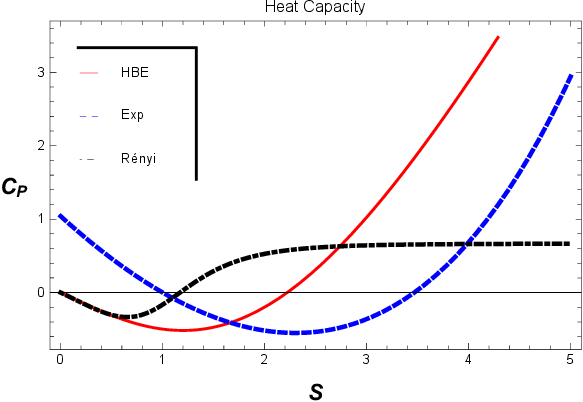,width=.45\linewidth} \caption{\raggedright $C_{P}$
versus the Hawking-Bekenstein, exponential corrected and R\'{e}nyi entropy $S$ by setting $P=1,~Q=1,~\alpha=1$ (in R\'{e}nyi entropy case). We obtain various curves for different entropy models to notice the impact of these entropies on $C_{P}$, such as the trajectories for the Hawking-Bekenstein, exponential corrected, and R\'{e}nyi entropies are demonstrated as red solid, blue dotted, and black dotted-dashed trajectories, respectively.}\label{Fig-1}
\end{figure}

Analyzing how specific heat varies with the size of a BH offers key insights into its stability under thermodynamic conditions and possible phase transitions. When the heat capacity is positive, the system tends to be stable and capable of equilibrating thermally with its surroundings; a negative value, on the other hand, points to instability, and a value of zero corresponds to a critical phase transition point. The heat capacity behavior in terms of different entropy frameworks is presented in Fig.~\ref{Fig-1} by setting $P=1,~\alpha=1$, and $Q=1$. Notably, we obtain different trajectories with respect to the entropy models that we employed in our analysis, such as for the Hawking-Bekenstein (red solid curve), exponential corrected entropy (blue dotted curve), and R\'{e}nyi entropy (black dotted-dashed curve). We want to mention here that in the legends of all the thermodynamic figures, we present the Hawking-Bekenstein and exponential corrected entropies by HBE and Exp, respectively. It observes that heat capacity in terms of Hawking-Bekenstein and R\'{e}nyi entropies initially shows negative behavior for small values of entropy, and then they show positive behavior as the entropy rises, and also for small values, we obtain ZP. For example, there is only one ZP or phase transition in the cases of the Hawking-Bekenstein (red solid curve) and R\'{e}nyi entropy (black dotted-dashed curve), which are at $2.2214$  and $1.1698$, respectively. However, in the case of exponential entropy (blue dotted curve), the heat capacity initially decreases with a positive behavior, and for small intervals of entropy, it becomes negative; however, after that, it again begins to grow with a positive behavior. In addition, we obtain two ZPs at $1$ and $3.4674$, indicating that in the presence of exponential entropy, we can obtain more ZPs compared to other entropy frameworks. In BH thermodynamics, ZPs act as pivotal points that distinguish between two unique thermal states; for example, they separate the unstable region, in which heat capacity shows negative behavior, from the stable region, in which heat capacity presents positive behavior. It is also often referred to as the restriction point in BH thermodynamic analysis, beyond which the system's behavior becomes stable. It is essential to mention here that by changing the parameter's value location of ZP may also change; for example, in our case, if one chooses different values of $Q,~P$ and $\alpha$ (in the case of R\'{e}nyi entropy), then the location of ZP might vary too. 
\subsection{$P-V$ Diagram and Isotherm compressibility}
In this subsection, we will discuss the $P-V$  and isothermal compressibility for a charged torus-like BH, which is also used to study the thermodynamic stability of the BH. First, we determine the equation of state using Eq.~\eqref{THB}, adopting the methodology described in Refs.~\cite{Kubiznak:2012wp,Kubiznak:2016qmn,Rodrigues:2022qdp,Rodrigues:2022zph}. In this way, one can obtain the equation of state in terms of conjugate temperature related to Hawking-Bekenstein entropy, Volume, and charge, which is given by
\begin{eqnarray}\label{PHB}
P(V_{\mathrm{HB}},~T_\mathrm{HB},~Q)&=&\frac{2 \times\ 2^{2/3} \left(\pi  Q^2+\frac{3 \pi  T V}{4}\right)}{3 \sqrt[3]{3 \pi } V^{4/3}}\,.
\end{eqnarray}
By adopting the same process that we opted for Hawking-Bekenstein entropy, but this time for the exponential corrected entropy by employing Eq.~\eqref{Tee}, which takes the following form
\begin{eqnarray}\label{Pee} P(V_{\mathrm{Exp}},~T_{\mathrm{Exp}},~Q)&=&\frac{8 (6 \pi )^{2/3} Q^2+9 \sqrt[3]{6} \pi ^{4/3} T \left(V^{4/3}\right)^{5/4}}{36 V^{4/3}}\,.\end{eqnarray}
In the context of R\'{e}nyi entropy, we utilize Eq.~\eqref{TRE} to compute the thermodynamic pressure in terms of the conjugate temperature, Volume, and charge, which is given as
\begin{eqnarray}\label{PRE} P(V_{\mathrm{R}},~T_{\mathrm{R}},~Q)&=&\frac{2 (2 \pi )^{2/3}}{3 \sqrt[3]{3} V^{4/3}} \left(Q^2+\frac{3 T \left(\alpha  V^{2/3}\right)^{3/2}}{\alpha ^{3/2} \left\{(6 \pi )^{2/3} \alpha  V^{2/3}+4\right\}}\right)\,.
\end{eqnarray}

\begin{figure} [t]
\epsfig{file=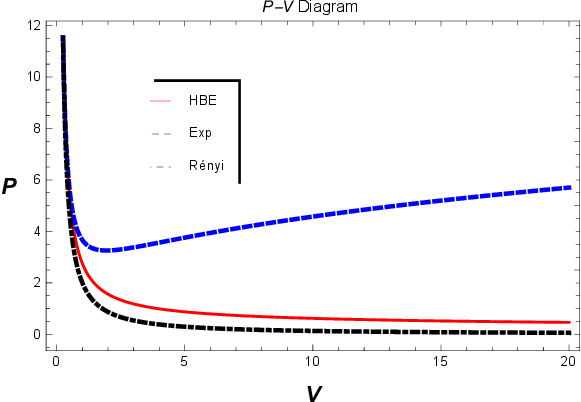,width=.45\linewidth} 
\epsfig{file=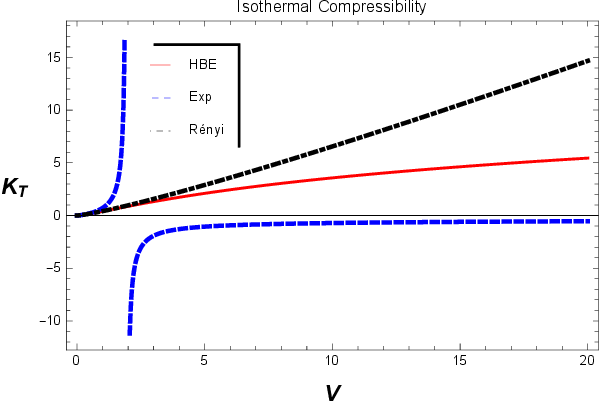,width=.45\linewidth} \caption{\raggedright $P-V$ (left panel) and isotherm compressbility $K_{T}$ (right panel)
in terms of the Hawking-Bekenstein, exponential corrected and R\'{e}nyi entropy by utilizing $P=1,~Q=1,~\alpha=1$ (in R\'{e}nyi entropy case). We get various curves for different entropy models to notice the effect of these entropies on $P-V$ (left panel) and $K_{T}$ (right panel), such as the trajectories for the Hawking-Bekenstein, exponential corrected, and R\'{e}nyi entropies are presented as red solid, blue dotted, and black dotted-dashed curves, respectively.}\label{Fig-2}
\end{figure}

In the context of BH systems, the impact of thermal fluctuations at equilibrium on the isothermal compressibility $(K_{T})$, which quantifies how the BH’s volume varies with pressure, is determined in Ref.~\cite{Rodrigues:2022zph}. It is directly computed by using the following formula
\begin{eqnarray}\label{ITC}
K_{T}=-\frac{1}{V_\mathrm{i}}\frac{\partial V_\mathrm{i}}{\partial P_\mathrm{i}}\bigg|_{T_\mathrm{i}}\,,
\end{eqnarray}
where $\mathrm{i}$ presents different entropy frameworks that we utilize in this thermodynamic analysis. By plugging Eqs.~\eqref{THB},~\eqref{Tee} and \eqref{TRE} into Eq.~\eqref{ITC}, it yields
\begin{eqnarray}\label{ICT1}
K_{T}(\mathrm{HB})&=&\frac{9 \sqrt[3]{6} V^{4/3}}{\pi ^{2/3} \left(16 Q^2+3 T V\right)}\,,\\\label{ICTC2} K_{T}(\mathrm{Exp})&=&\frac{18 \left(\frac{6}{\pi }\right)^{2/3} V^{4/3}}{32 \sqrt[3]{6} Q^2-9 \pi ^{2/3} T \left(V^{4/3}\right)^{5/4}}\,,\\\nonumber  K_{T}(\mathrm{R})&=&\left[9 \sqrt[3]{3} \sqrt{\alpha } V^{4/3} \left((6 \pi )^{2/3} \alpha  V^{2/3}+4\right)^2\right]\bigg[2 (2 \pi )^{2/3} \bigg\{8 \sqrt{\alpha } Q^2 \bigg(3 \sqrt[3]{6} \pi ^{4/3} \alpha ^2 V^{4/3}\\\label{ICTC3}&+&4 (6 \pi )^{2/3} \alpha  V^{2/3}+8\bigg)+\frac{3 T \left(3 (6 \pi )^{2/3} \alpha  V^{2/3}+4\right) \left(\alpha  V^{2/3}\right)^{3/2}}{\alpha }\bigg\}\bigg]^{-1}\,.
\end{eqnarray}

$P-V$ diagram is considered to be one of the methods to analyze the stability of the BH system. In Fig.~\ref{Fig-2} left panel, we depict the behavior of $P-V$ diagram by utilizing $Q=1,~P=1$ and $\alpha=1$ while we obtain different trajectories in terms of respective entropy models such as the Hawking-Bekenstein (red solid curve), exponential corrected entropy (blue dotted curve) and the R\'{e}nyi entropy (black dotted-dashed curve). It observes that pressure $(P)$ smoothly decreases as the volume $(V)$ grows for the Hawking-Bekenstein (red solid curve) and R\'{e}nyi entropy (black dotted-dashed curve). This smooth behavior indicates the absence of a possible first-order phase transition, such as the small and large BH transitions corresponding to the Van-dar-Waals fluctuation in the $P-V$ diagram. Therefore, this signifies that the BH remains thermodynamically stable, undergoing no abrupt transitions in its internal configuration. However, in the context of exponential corrected entropy (blue dotted curve), we observe that initially, pressure decreases for small values of volume. The pressure begins to increase as the volume increases, which indicates a possible phase transition in the thermodynamic system. In the right panel of Fig.~\ref{Fig-2}, we demonstrate the isotherm compressibility $(K_{T})$ in terms of the Hawking-Bekenstein (red solid curve), exponential corrected entropy (blue dotted curve), and R\'{e}nyi entropies (black dotted-dashed curve) by plugging $T=1,~Q=1$ and $\alpha=1$ (R\'{e}nyi entropy case). As we mentioned earlier in the $P-V$ diagram (left panel), it also validates from the behavior of isotherm compressibility $(K_{T})$ in the form of exponential corrected entropy (blue dotted curve) that there is a possible phase transition while in cases of the Hawking-Bekenstein entropy (red solid curve) and the R\'{e}nyi entropy (black dotted-dashed curve) shows increasing behavior for all the values of volume.  
\subsection{Helmholtz Free Energy}
In this subsection, we will discuss the global stability of the charged torus-like BH, which is associated with the Helmholtz and Gibbs free energy. Firstly, we investigate the Helmholtz free energy by using our concerned entropy frameworks. As described in Refs.~\cite{NaveenaKumara:2020biu,Rani:2025esb,Rodrigues:2022zph,Dehyadegari:2017flm}, the Helmholtz free energy is utilized to assess the global stability of BHs and is defined as follows
\begin{eqnarray}\label{hfe}
\mathcal{F}&=&H-T_\mathrm{i}S_\mathrm{i}\,,
\end{eqnarray}
where $H$ represents the enthalpy of the system. In extended thermodynamics, the mass of a BH represents the system's enthalpy rather than its internal energy. Therefore, it is quite easy to obtain HFE in terms of Hawking-Bekenstein entropy by inserting Eqs.~\eqref{MHB} and \eqref{THB} in Eq.~\eqref{hfe}, which takes the following form
\begin{eqnarray}\label{hfeHB}
\mathcal{F}(S_\mathrm{HB})&=&\frac{3 \pi  Q^2}{\sqrt{S}}-\frac{2 P S^{3/2}}{3 \pi }\,.
\end{eqnarray}
In the case of exponential corrected entropy, we derive HFE by employing Eqs.~\eqref{Mee} and \eqref{Tee} in Eq.~\eqref{hfe}. It is expression is presented as 
\begin{eqnarray}\label{Cee} \mathcal{F}(S_\mathrm{Exp})&=&\frac{4 P (S-4) (S-1)+3 \pi ^2 Q^2 (5 S-4)}{6 \sqrt[4]{2} \pi  (S-1)^{5/4}}\,.
\end{eqnarray}
Similarly, one can obtain the HFE by putting Eqs.~\eqref{MRE} and \eqref{TRE} in Eq.~\eqref{hfe}. Its expression is given as 
\begin{eqnarray}\label{CRE} \mathcal{F}(S_\mathrm{R})&=&\frac{\pi  \sqrt{\alpha } Q^2 \left(e^{\alpha  S} (\alpha  S+2)-2\right)}{\left(e^{\alpha  S}-1\right)^{3/2}}-\frac{2 P \sqrt{e^{\alpha  S}-1} \left(e^{\alpha  S} (3 \alpha  S-2)+2\right)}{3 \pi  \alpha ^{3/2}}\,.
\end{eqnarray}

\begin{figure}[t]
\epsfig{file=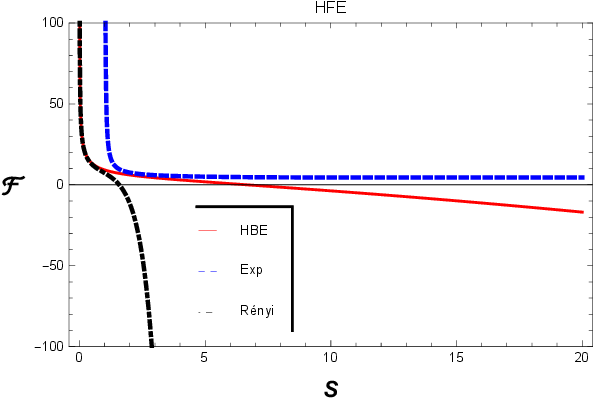,width=.45\linewidth}  \caption{\raggedright HFE in terms of the Hawking-Bekenstein, exponential corrected, and R\'{e}nyi entropy by substituting $P=1,~Q=1,~\alpha=1$ (in R\'{e}nyi entropy case). We get various curves for different entropy models to observe the effect of these entropies on HFE, such as the trajectories for the Hawking-Bekenstein, exponential corrected, and R\'{e}nyi entropies are depicted as red solid, blue dotted, and black dotted-dashed curves, respectively.}\label{Fig-3}
\end{figure}

We investigate HFE for the charged torus-like BHs in terms of the Hawking-Bekenstein, exponential corrected, and R\'{e}nyi entropies, which is considered one of the methods associated with BHs' global stability. Behavior of HFE as the function of Hawking-Bekenstein entropy (red solid curve), exponential corrected (blue dotted curve) and R\'{e}nyi entropy (black dotted-dashed curve) by setting $Q=1,~P=1$ and $\alpha=1$ is depicted in Fig.~\ref{Fig-3}. It shows that initially HFE decreases with positive behavior, but as entropy increases, the behavior of HFE decreases with negative behavior for both the Hawking-Bekenstein and R\'{e}nyi entropies. This fluctuating behavior suggests that charged torus-like BHs are not stable in terms of the Hawking-Bekenstein (red solid curve) and R\'{e}nyi (black dotted-dashed curve) entropies. Surprisingly, HFE shows positive behavior for all the ranges of entropy, which suggests stability of HFE for the charged torus-like BHs. Furthermore, this indicates that the thermodynamic behavior of BHs changes based on the entropy formalism used, with both Hawking-Bekenstein and Rényi entropies becoming unstable under specific conditions, such as varying charge or cosmological constant, suggesting limitations in their applicability within particular physical scenarios.  
\subsection{Gibbs Free Energy}
In this subsection, we discuss the Gibbs free energy of the charged torus-like BH, which is also related to global stability. As described in Refs.~\cite{Kubiznak:2012wp,Kubiznak:2016qmn,Rani:2025esb,Rodrigues:2022zph}, the Gibbs free energy is utilized to assess the global stability of BHs and is defined as follows
\begin{eqnarray}\label{gfe}
\mathcal{G}&=&H-T_\mathrm{i}S_\mathrm{i}+P_\mathrm{i}V_\mathrm{i}\,.
\end{eqnarray}
By plugging  Eqs.~\eqref{MHB}, \eqref{THB} in Eq.~\eqref{gfe}, one can calculate the Gibbs free energy in terms of the Hawking-Bekenstein, and it takes the following shape
\begin{eqnarray}\label{gfeHB}
\mathcal{G}(S_\mathrm{HB})&=&\frac{6 P S^2+17 \pi ^2 Q^2}{4 \pi ^2 Q^2 S-8 P S^3}\,.
\end{eqnarray}
By following the same procedure that we mentioned above, we determine GFE by utilizing Eqs.~\eqref{Mee} and \eqref{Tee} in Eq.~\eqref{gfe}, which can be written as
\begin{eqnarray}\label{gfeee} \mathcal{G}(S_\mathrm{Exp})&=&\frac{4 P (S-1) (5 S-8)+3 \pi ^2 Q^2 (5 S-4)}{6 \sqrt[4]{2} \pi  (S-1)^{5/4}}\,.
\end{eqnarray}
Similarly, by inserting Eqs.~\eqref{MRE},~\eqref{TRE} in Eq.~\eqref{gfe}, we calculate GFE in the form of R\'{e}nyi entropy, which is given as
\begin{eqnarray}\label{gfeRE}
\mathcal{G}(S_\mathrm{R})&=&\frac{\pi  \sqrt{\alpha } Q^2 \left(e^{\alpha  S} (\alpha  S+2)-2\right)}{\left(e^{\alpha  S}-1\right)^{3/2}}-\frac{2 P \sqrt{e^{\alpha  S}-1} \left(e^{\alpha  S} (3 \alpha  S-4)+4\right)}{3 \pi  \alpha ^{3/2}}\,.
\end{eqnarray}
\begin{figure} [t]
\epsfig{file=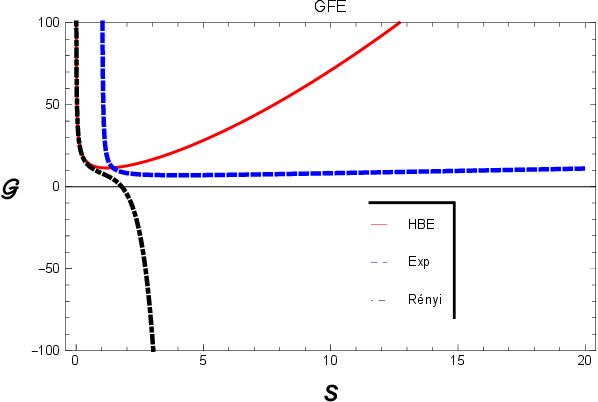,width=.45\linewidth}  \caption{\raggedright GFE in terms of the Hawking-Bekenstein, exponential corrected, and R\'{e}nyi entropy by utilizing $P=1,~Q=1,~\alpha=1$ (in R\'{e}nyi entropy case). We get various curves for different entropy models to notice the impact of these entropies on GFE, such as the trajectories for the Hawking-Bekenstein, exponential corrected, and R\'{e}nyi entropies are demonstrated as red solid, blue dotted, and black dotted-dashed curves, respectively.}\label{Fig-4}
\end{figure}

In  Fig.~\ref{Fig-4}, we present the graphical behavior of GFE as a function of the Hawking-Bekenstein, exponential corrected entropy, and R\'{e}nyi entropies by putting $Q=1,~p=1$ and $\alpha=1$. One can notice that initially, GFE in terms of the R\'{e}nyi entropy (black dotted-dashed curve) decreases with positive behavior, but as the entropy increases, GFE continuously decreases with a negative trend, which indicates that in the R\'{e}nyi entropy case, the behavior of GFE is not stable. However, for both the Hawking-Bekenstein (red solid curve) and the exponential-corrected (blue dotted curve) entropies, GFE initially decreases, but as entropy increases, it begins to rise, indicating its stability across all entropy ranges. 

Furthermore, the comparative analysis of all the above-mentioned thermodynamic quantities reveals that the exponential corrected entropy offers a more intricate phase structure, characterized by two specific heat ZPs absent in the Hawking-Bekenstein and R\'{e}nyi cases. This distinction reveals a more profound dependence of the microstructure on the horizon's topology, which we explore geometrically in the following section by employing thermodynamic curvature.  

\section{Thermodynamic Geometry in terms of Various entropy frameworks}\label{TDG}
This section presents different thermodynamic geometry formalisms (specifically, the Weinhold and Ruppiener metrics) based on various entropy corrections and examines how the divergence points align with ZP of the heat capacity \cite{Soroushfar:2020wch}. The application of geometric frameworks through thermodynamic geometry has notably advanced our understanding of BH thermodynamic structures. In the context of thermal fluctuation theory, thermodynamic geometry provides an effective method for exploring the microscopic nature and phase transitions of BHs.
We mention here that the curvature scalar, defined as an invariant over the thermodynamic space, provides essential insight into microscopic interactions and indicates critical behavior. This framework is frequently employed to study the nature of interactions within BH microstructures. The geothermodynamics methodology is utilized in this work to investigate the detailed connections between thermodynamic parameters and the corresponding geometric framework. We validate the phase transition for charged torus-like BHs by using different entropy frameworks. To begin with, we first define the basic framework of the thermodynamic geometry formalism. Therefore, we adopt the same process as described in  Ref.~\cite{Weinhold:1975xej}, and the Weinhold metric framework is defined as   
\begin{eqnarray}\label{6be}
g_{\nu\omega}^\mathrm{W}=\partial_{\nu}\partial_{\omega} M(S,~P,~Q)\,,
\end{eqnarray}
the line element of the Weinhold metric for torus-like BH is
\begin{eqnarray}\label{7be}
dS^{2}_\mathrm{W}=M_{SS}dS^{2}+M_{PP}dP^{2}+M_{QQ}dQ^{2}+2M_{SP}dSdP+2M_{PQ}dPdQ+2M_{SQ}dSdQ\,,
\end{eqnarray}
and the corresponding metric according to Eq.~\eqref{7be} can be written as\begin{equation}\label{M11}
g^\mathrm{W}=
\begin{pmatrix}
M_{SS} & M_{SP} & M_{PQ}  \\
M_{PS} & M_{PP} & 0 \\ M_{QS} & 0
  & M_{QQ} \\
\end{pmatrix}\,.
\end{equation}

Here, we mention that the Ricci curvature scalar obtained from the Weinhold metric formalism in terms of the Hawking-Bekenstein, exponential corrected and R\'{e}nyi entropies are zero (this signifies that the equilibrium manifold is flat in this metric formulation) which allows us to adopt another metric formalism that we employ in this study, is the Ruppeiner metric to investigate the coincidence of the divergence of the Ricci curvature scalar with the ZPs of heat capacity. In this metric formalism, we basically modify the Weinnhold metric given in Eq.~\eqref{MM1} by multiplying it with the inverse temperature. The purpose of this scaling, from a physical perspective, is to ensure the consistency of the fluctuation theory underlying the Ruppeiner metric with temperature-dependent thermodynamic probability distributions. For example, the inclusion of inverse temperature enhances our understanding of curvature singularities as indicators of critical behavior or phase transitions in the context of BH thermodynamics. Therefore, the inverse temperature serves not just as a mathematical term but as a thermodynamically significant quantity that influences the structure of fluctuation geometry and reveals key aspects of stability and microscopic behavior. Thereby, one can define the Ruppeiner metric formalism given in Refs.~\cite{Ruppeiner:2008kd,Ruppeiner:1995zz,Akbar:2011qw,Soroushfar:2020wch}, whose explicit form is given as 
\begin{eqnarray}\label{9be}
dS^{2}_\mathrm{Rup}=1/T(dS^{2}_{W})\,,
\end{eqnarray}
and the corresponding metric is
\begin{equation}\label{MM1}
g^\mathrm{Rup} =\frac{1}{T}
\begin{pmatrix}
M_{SS} & M_{SP} & M_{PQ}  \\
M_{PS} & M_{PP} & 0 \\ M_{QS} & 0
  & M_{QQ} \\
\end{pmatrix}\,.
\end{equation}

Firstly, we obtain the curvature scalar from the Ruppeiner metric formalisms in terms of Hawking-Bekenstein entropy by plugging Eqs.~\eqref{MBH} and \eqref{THB} into  Eq.~\eqref{MM1}, which yields
\begin{eqnarray}\label{RRSHB}
R^\mathrm{Rup}(S_\mathrm{HB})=\frac{6 P S^2+17 Q^2}{4 Q^2 S-8 P S^3}\,.
\end{eqnarray}

In addition, we obtain the Ricci curvature scalar from the Ruppeiner formalism in the form of exponential corrected entropy by inserting Eqs.~\eqref{Mee} and \eqref{Tee} into Eq.~\eqref{MM1}, and it is given as
\begin{eqnarray}\label{RRSee}
R^\mathrm{Rup}(S_\mathrm{Exp})=-\frac{-4 P (S-1)-19 \pi ^2 Q^2}{8 (S-1) \left(\pi ^2 Q^2-4 P (S-1)\right)}\,.
\end{eqnarray}

Similarly, one can easily derive the Ricci curvature of the Ruppeiner formalism in the presence of R\'{e}nyi entropy by substituting Eqs.~\eqref{MRE} and \eqref{TRE} into Eq.~\eqref{MM1}, which takes the following shape
\begin{eqnarray}\label{RRSRE}
R^\mathrm{Rup}(S_\mathrm{R})=-\frac{\alpha  \left\{2 P \left(5 e^{\alpha  S}-2\right) \left(e^{\alpha  S}-1\right)^2+\pi ^2 \alpha ^2 Q^2 \left(15 e^{\alpha  S}+2\right)\right\}}{4 \left(e^{\alpha  S}-1\right) \left\{2 P \left(e^{\alpha  S}-1\right)^2-\pi ^2 \alpha ^2 Q^2\right\}}\,.
\end{eqnarray}

\begin{figure} \centering
\epsfig{file=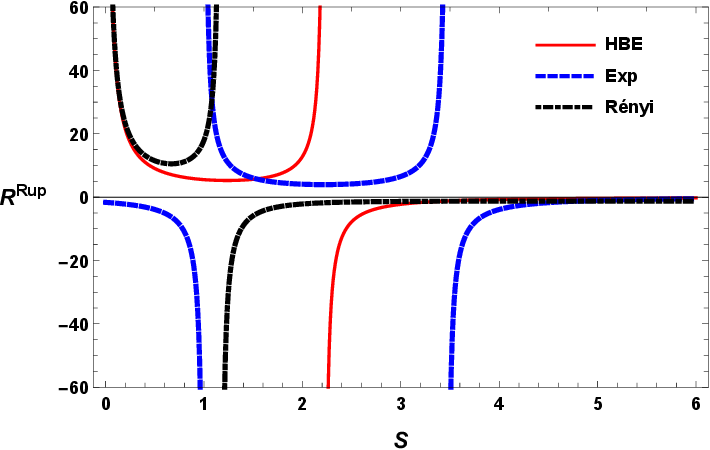,width=.45\linewidth}  \caption{\raggedright Ricci curvature in terms of the Hawking-Bekenstein, exponential corrected, and R\'{e}nyi entropies by inserting $P=1,~Q=1,~\alpha=1$ (in R\'{e}nyi entropy case). We obtain different curves for different entropy models to witness the impact of these entropies on curvature scalar $R^\mathrm{Rup}$, such as the trajectories for the Hawking-Bekenstein, exponential corrected, and R\'{e}nyi entropies are illustrated as red solid, blue dotted, and black dotted-dashed curves, respectively.}\label{Fig-5}
\end{figure}

One can obtain valuable insights regarding the possible phase transition and the microscopic interaction between the particles of the BHs by examining the Ricci curvature scalar. For example, a negative scalar curvature indicates strong attractive microscopic interactions, whereas a positive value suggests repulsive interactions between the particles. When the curvature is flat, it typically denotes either an ideal gas-like non-interacting system or a perfectly balanced interaction regime. The behavior of the curvature scalar obtained from the aforementioned formalism in various entropy frameworks by substituting $Q=1,~P=1$ and $\alpha=1$ is presented in Fig.~\ref{Fig-5}. The trajectories that we plot in Fig.~\ref{Fig-5} are the Ricci scalar in terms of entropies such as the Hawking-Bekenstein (red solid curve),~exponential corrected (blue dotted curve), and R\'{e}nyi entropies (black dotted-dashed curve). It observes that the ZPs of heat capacity also coincide with the divergence of the Ricci curvature scalar. In the Hawking-Bekenstein and R\'{e}nyi entropies cases, the Ricci curvature scalar $R^\mathrm{Rup}$ divergence coincides with ZP of heat capacity at $2.2214$  and $1.1698$, respectively. Similarly, in the case of the exponential corrected entropy, the divergence of $R^\mathrm{Rup}$ coincides with both ZPs of heat capacity at $1$ and $3.4674$. 

One can also notice that the behavior of the Ricci curvature scalar is both negative and positive, which indicates that there is both attractive and repulsive interaction between the particles of the charged torus-like BHs that can be related to the non-trivial toroidal topology. Owing to the dual periodic identification of the $S^{*}\times S^{*}$ horizon, the fluctuation manifold gains an additional curvature freedom, enabling $R^\mathrm{Rup}$ to be highly sensitive to small changes in entropy and providing a dual transitional behavior. In a broader context, the singularities in $R^\mathrm{Rup}$ correspond to topological defects within the thermodynamic state manifold, where their numbers of divergence (or ZP) corresponds to the phase transition order; in our case (toroidal BH), the existance of two such divergence in $R^\mathrm{Rup}$ signifies a modification in the thermodynamic manifold's topological charge, which also agrees with the recent studies on thermodynamic topology \cite{Wei:2020poh,Bhattacharya:2019qxe,Wei:2021vdx,Jeon:2024yey}. The Alignment of divergence points of $R^\mathrm{Rup}$ with the ZP of heat capacity confirms that geothermodynamic formalisms serve as a consistent and independent tool for diagnosing phase transitions.      

\subsection{A detailed thermodynamic comparison of various entropy models}
In this subsection, we compare the thermodynamic behavior obtained from the different entropy models along with the classical BH solutions, such as Schwarzschild and RN-AdS BHs. For completeness, we employ the generalized three-parameter entropy (GTPE) model and compare it with the entropy models we discussed in this study. Firstly, we discuss the mathematical expression of GTPE, which is obtained from the generalised entropy frameworks given in Refs.~\cite{Elizalde:2025iku,Nojiri:2022aof,Nojiri:2024zdu}, and its expression can be written as 
\begin{eqnarray}\label{Gtpe}
    \mathcal{S}_{3}[\alpha,~\beta,~\gamma]=\frac{1}{\gamma}\left\{\left(1+\frac{S_\mathrm{HB}\alpha}{\beta}\right)^{\beta}-1\right\}\,,
\end{eqnarray}
where $\alpha,~\beta,~\gamma\geq0$ are the free parameters. We mention here that this entropy is the reduced form of generalized entropy given in Refs.~\cite{Nojiri:2024zdu}, which contains six free parameters. Now, if one can choose $\gamma=\alpha$, one can obtain the Sharma-Mittal entropy \cite{jahromi2018generalized,Nojiri:2024zdu}, while if one chooses $\alpha\to\infty$, then Eq.~\eqref{Gtpe} leads to the Tsallis and Barrow entropy \cite{tsallis1988possible,Barrow:2020tzx}. Similarly, by choosing  $\alpha,~\beta\to\infty$ with $\frac{\alpha}{\beta}$ finite, which eventually yields the Rényi entropy \cite{Odintsov:2022qnn}, and also if we put $\gamma=\alpha,~\beta\to\infty$, then we can obtain the Loop Quantum gravity entropy \cite{Majhi:2017zao}. Additionally, one can easily recover the Hawking-Bekenstein entropy \cite{Bekenstein:1973ur,Nojiri:2024zdu}, by replacing $\alpha_{+}\to\infty,~\beta=1,~\gamma=(\alpha/\beta)^{\beta}$ in Eq.~\eqref{Gtpe}. 

\begin{table}[t]
\caption{\label{tab:comparison}\raggedright This table presents the thermodynamic quantities by employing the Hawking-Bekenstein (HB), R\'{e}nyi, and exponential corrected (EC) entropies for the charged torus-like BH, and also its comparison with the thermodynamic quantities of the RN-Ads and Schwarzschild BHs. For a comprehensive analysis, we compared our entropy models with the generalized entropy by incorporating GTPE from Refs.~\cite{Elizalde:2025iku,Nojiri:2022aof}. Here, we present the thermodynamic quantities like temperature, heat capacity, isothermal compressibility, HFE, GFE, and  Ruppeiner Ricci scalar by $C,~K_{T}~\mathcal{F},~\mathcal{G}$, and $R^\mathrm{Rup}$, respectively.}
\begin{ruledtabular}
\small
\begin{tabular}{ccccccc}
\textbf{BHs}  &  &     &  \textbf{CTL BH}  &  & \textbf{RN-AdS} & \textbf{Schwarzschild}\\ 
\hline
\textbf{Quantitites}   &\textbf{HBE} & \textbf{Rényi} & \textbf{ECE}& \textbf{GTPE} & \textbf{HBE} & \textbf{HBE}\\
\hline
$C$         & UnStable        & Unstable & UnStable  &  UnStable&Unstable& Unstable                \\
ZPs          & 1              & 1               & 2 &1 & 1  & 0             \\
$P-V$          & Monotonic              & Monotonic               & Non-monotonic & Monotonic   & Non-monotonic &  not defined \\
$K_{T}$       & Positive                & Positive               & Divergence & Positive  & Divergence &   not defined           \\
$\mathcal{F}$  & UnStable               & Unstable                & Stable & Unstable &Unstable &Unstable
             \\
$\mathcal{G}$  & Stable               & Unstable                & Stable& Unstable&Stable & Stable           \\
$R^\mathrm{Rup}$  & Coincide               & Coincide                & Coincide both ZPs & Coincide   & Coincide&   No-coincidence    
\\
\end{tabular}
\end{ruledtabular}
\end{table}

Table~\ref{tab:comparison} summarizes the thermodynamic quantities for the charged torus-like BHs derived employing the Hawking-Bekenstein, R\'{e}nyi, exponential corrected, and the GTPE model, together with those for RN-AdS and Schwarzschild BHs. It becomes from our analysis that the thermodynamic behavior of the Hawking-Bekenstein, R\'{e}nyi, and GTPE models is qualitatively similar. In contrast, the exponentially corrected entropy yields a distinct trend in all thermodynamic quantities. For example, exponential corrected entropy leads to two heat capacity ZPs, while the remaining entropy formulations give only one ZP, a behavior further confirmed by the associated Ruppeiner curvature divergences. This comparison indicates that the exponential corrected entropy responds more sensitively to BH microstructure and horizon topology.

\section{Sparsity of Hawking radiations and Emission of Energy through Different Entropy Frameworks}\label{SHR}
The purpose of this section is to delve into BH sparsity, demonstrating that BHs emit radiation akin to blackbody radiation, where their temperature is set by the surface gravity. The evaporation of a BH gives rise to Hawking radiation, which is notably sparse and distinctly different from classical blackbody emission. Sparsity, as described in Refs.~\cite{Page:2004xp,Gray:2015pma}, reflects the average spacing between emitted quanta, a quantity that depends on the energies involved in the emissions. Its explicit expression is given as 

\begin{eqnarray}\label{SP}
    \eta =\frac{\mathbb{C}}{\mathbb{g} }\left(\frac{\lambda_t^2}{\mathcal{A}_\mathrm{eff}}\right)\,,
\end{eqnarray}
where $\mathbb{C}$ represents the dimensionless constant  thermal wavelength is denoted by $\lambda_{\mathfrak{t}}$, the effective BH area is $\mathcal{A}_\mathrm{eff}=27\mathcal{A}_\mathrm{BH}/4$ and $\mathbb{g}$ denotes the spin degeneracy of the emitted particles. In the context of a simple Schwarzschild BH radiating massless spin-1 particles, the value $\eta_\mathrm{Sh}\approx 73.49$ is used (for further details, check Ref.~\cite{Gray:2015pma,Rani:2025esb}). It is worth recalling, for comparison, that $\eta$ is significantly smaller than 1 in the case of black body radiation.
\begin{eqnarray}\nonumber
    \eta_\mathrm{HB} &=&\frac{4\mathbb{C} \pi ^3 S^2}{27\mathbb{g} \left(Q^2-2 P S^2\right)^2}~,\\\nonumber
    \eta_\mathrm{Exp} &=&\frac{16 \mathbb{C} \pi ^4 (S-1)^2}{27 \mathbb{g}\left(\pi ^2 Q^2-4 P (S-1)\right)^2}~,\\\label{SP1}
    \eta_\mathrm{R} &=&\frac{4 \mathbb{C} \pi ^4 \alpha ^2 e^{-2 \alpha  S} \left(e^{\alpha  S}-1\right)^2}{27\mathbb{g} \left(\pi ^2 \alpha ^2 Q^2-2 P \left(e^{\alpha  S}-1\right)^2\right)^2}\,.
\end{eqnarray}
\begin{figure}[t]
     \begin{subfigure}{0.35\textwidth}
         \includegraphics[width=\textwidth]{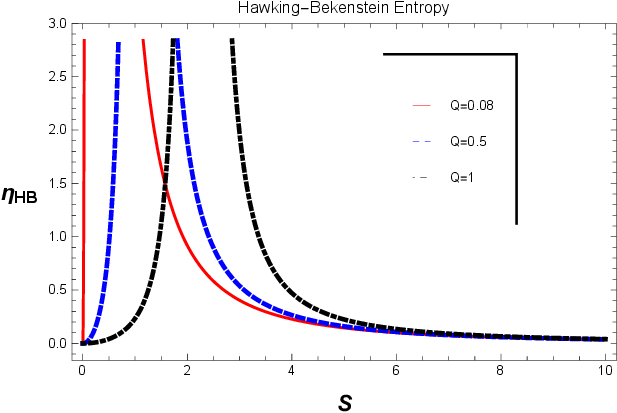}
         \caption{Hawking-Bekenstein entropy.}
         \label{SPHB}
     \end{subfigure}
     \begin{subfigure}{0.35\textwidth}
         \includegraphics[width=\textwidth]{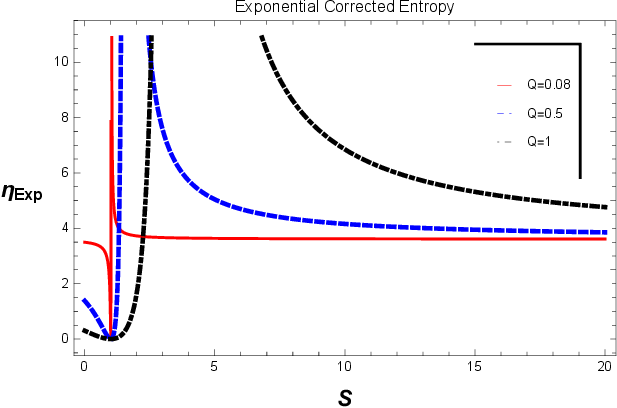}
         \caption{Exponential entropy.}
         \label{SPexp}
     \end{subfigure}
     \begin{subfigure}{0.35\textwidth}
         \includegraphics[width=\textwidth]{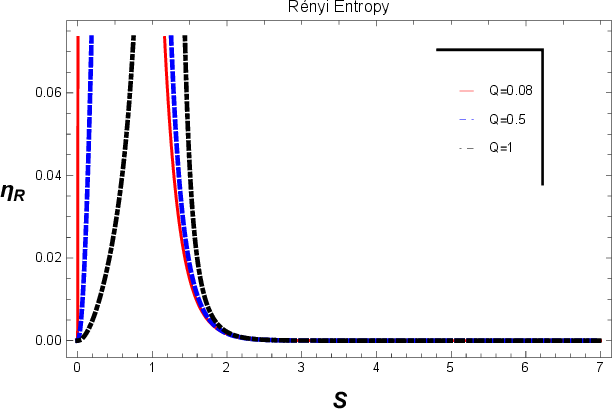}
         \caption{R\'{e}nyi entropy with $Q$ variations.}
         \label{SPR}
     \end{subfigure}
     \begin{subfigure}{0.35\textwidth}
         \includegraphics[width=\textwidth]{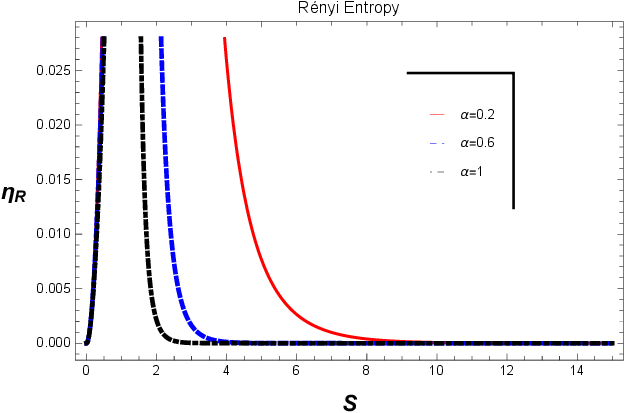}
         \caption{R\'{e}nyi entropy with $\alpha$ variations.}
         \label{SPP}
     \end{subfigure}
        \caption{\raggedright Sparsity $\eta$ as the function of different entropies such as the Hawking-Bekenstein, exponential corrected, and R\'{e}nyi entropies by setting $P=1,~\alpha=1$ (in R\'{e}ny entropy case) and we substitute $Q=0.08$ (red solid curve), $Q=0.5$ (blue dotted curve) and $Q=1$ (black dotted-dashed curve) for different curves, in Figs.~(\subref{SPHB})-(\subref{SPR}). However, in Fig.~(\subref{SPP}), we insert $Q=1,~P=1$ and for red solid, blue dotted, and black dotted-dashed curves we utilize $\alpha=0.2,~0.6,~1$, respectively.}
        \label{Fig-6}
\end{figure}

Let us mention here that in this analysis of sparsity, we employed the conjugate temperature corresponding to the entropy framework to maintain consistency. To analyze the influence of BH system parameters on Hawking sparsity using different entropy approaches, Fig.~\ref{Fig-6} presents $\eta$ as a function of entropy $S$ relationship at  $P=1$ and different trajectories corresponding to charge $Q=0.08$ (red solid curve), $Q=0.5$ (blue dotted curve), and $1$ (black dotted-dashed curve). We present the sparsity of Hawking radiation by using Hawking-Bekenstein entropy in Fig.~\ref{Fig-6}(\subref{SPHB}) and exponential corrected entropy in Fig.~\ref{Fig-6}(\subref{SPexp}). Similarly, we present $\eta$ as the function of entropy by putting $Q=0.08,~0.5,~1$ for R\'{e}nyi entropy in Fig.~\ref{Fig-6}(\subref{SPR}) while in  Fig.~\ref{Fig-6}(\subref{SPP}) we employ different values of non-extensive parameter $\alpha=0.2,~0.6,~1$. It observes that as entropy increases, it leads to a reduction in sparsity, which is not the case for Schwarzschild BHs. Also, for smaller values of entropy, sparsity $\eta$ increases more than the Schwarzschild one, which indicates that, during this phase of evaporation, the BH emits radiation with a lower intensity than Hawking radiation. As 
$S$ increases, $\eta$ steadily decreases, and approaches zero in the limit. In the case of exponential corrected entropy, $\eta$ decreases as the entropy increases, but after some values of $S$, it shows constant behavior, which is quite evident in Fig.~\ref{Fig-6}(\subref{SPexp}). Significant changes in the decay behavior arise either due to modifications in the BH parameter space or from adopting different entropy formalisms.

The interior of the BH experiences intense quantum fluctuations that cause particle annihilation and birth apart from its event horizon.  The tunneling mechanism plays a significant role in the evaporation of BHs by enforcing the particles with positive charge from the innermost region of Hawking radiation, beyond the horizon of the BH, during a particular time period. BH evaporation is intrinsically tied to the rate of energy emission, with the cross-section for high-energy absorption, as perceived by a distant observer, approximating the BH's shadow. The cross-section for energy absorption by the BHs varies with respect to the fixed value of $\sigma_{lim}$. By adopting a similar methodology mentioned in Refs.~\cite{Wei:2013kza,Ditta:2022fjz}, one can express this constant value that resolves the radius of the BH as 
\begin{eqnarray}\label{sig}
    \sigma_\mathrm{lim} &\approx&\pi r^{2}_\mathrm{e}\,.
\end{eqnarray}
Thereby, one can express the relation for the rate of energy emission of BHs, given as 
\begin{eqnarray}\label{sig}
    \frac{d^{2}\varepsilon}{dt d\omega } &=&\frac{2\sigma_\mathrm{lim}\pi^{2}\omega^{3}}{e^{\omega/T}-1}\,,
\end{eqnarray}
\begin{figure}[t]
     \begin{subfigure}{0.35\textwidth}
         \includegraphics[width=\textwidth]{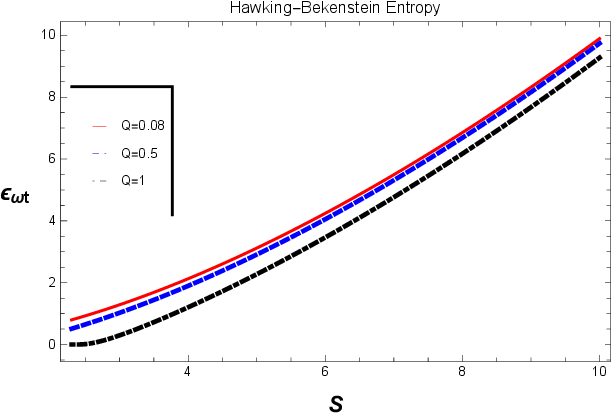}
         \caption{Hawking-Bekenstein entropy.}
         \label{EEHB}
     \end{subfigure}
     \begin{subfigure}{0.35\textwidth}
         \includegraphics[width=\textwidth]{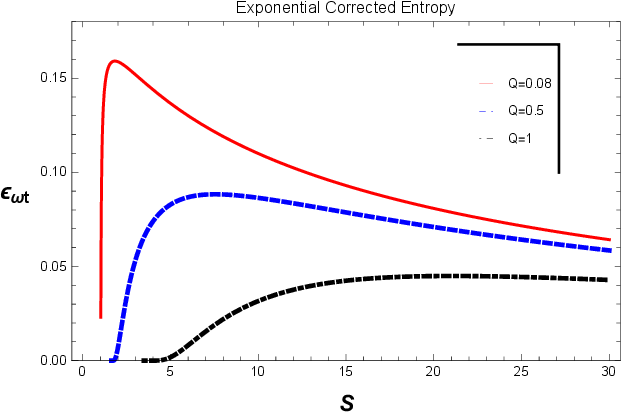}
         \caption{Exponential entropy.}
         \label{EEexp}
     \end{subfigure}
     \begin{subfigure}{0.35\textwidth}
         \includegraphics[width=\textwidth]{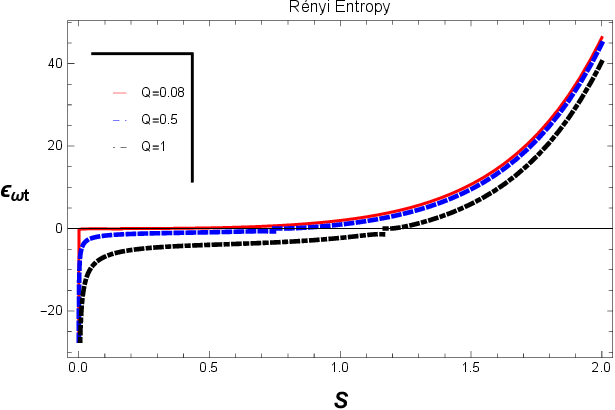}
         \caption{R\'{e}nyi entropy with $Q$ variations.}
         \label{EER}
     \end{subfigure}
     \begin{subfigure}{0.35\textwidth}
         \includegraphics[width=\textwidth]{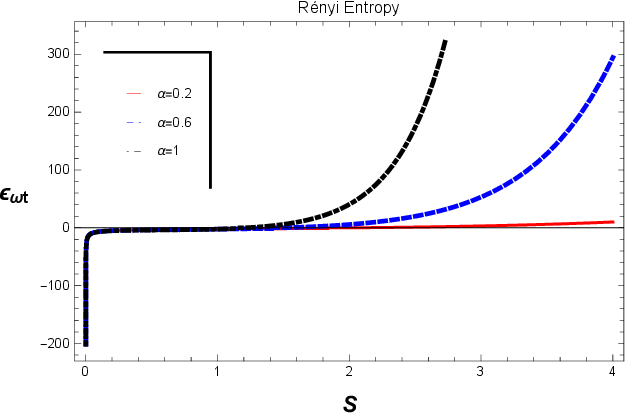}
         \caption{R\'{e}nyi entropy with $\alpha$ variations.}
         \label{EEE}
     \end{subfigure}
        \caption{\raggedright The rate of energy emission $\varepsilon_{\omega t}$ as the function of different entropies such as the Hawking-Bekenstein, exponential corrected, and R\'{e}nyi entropies by setting $P=1,~\omega=1,~\alpha=1$ (in R\`{e}ny entropy case) and for different curves we put $Q=0.08$ (red solid curve), $Q=0.5$ (blue dotted curve) and $Q=1$ (black dotted-dashed curve), in Figs.~(\subref{EEHB})-(\subref{EER}). But for Fig.~(\subref{EEE}), we put $Q=1,~P=1$ and $\alpha=0.01,~0.5,~1$ for red solid, blue dotted, and black dotted-dashed curves, respectively.}
        \label{Fig-7}
\end{figure}where $T$ is the temperature, which in our analysis corresponds to the different entropy frameworks such as the Hawking-Bekenstein,  exponential corrected, and R\'{e}nyi entropies. We plot the rate of energy emission $\varepsilon_{\omega t}$ in the form of the Hawking-Bekenstein, exponential corrected, and R\'{e}nyi entropies in Fig.~\ref{Fig-7} by setting $\alpha=1$ and we obtain different trajectories for  $Q=0.08$ (red solid curve), $Q=0.5$ (blue dotted curve), and $Q=1$ (black dotted-dashed curve) for Figs.~\ref{Fig-7}(\subref{EEHB}-\subref{EER}) while in Fig.~\ref{Fig-7}(\subref{EEE}), we put $Q=1,~\omega=1,~P=1$ and $\alpha=0.2$ (red solid curve), $\alpha=0.6$ (blue dotted curve) and $\alpha=1$ (  black dotted-dashed dashed curve).  For a charged torus-like BH,  it is noticed that the rate of energy emission is increasing initially with a negative trend for small ranges of entropy, but as the entropy increases, the values of $\varepsilon_{\omega t}$ also increase with a positive trend. While in the case of exponential corrected entropy, we observe that the rate of emission energy is increasing for small values of entropy with a positive trend, and for large values of entropy, the rate of emission energy slightly decreases, and after that, it becomes constant, which we can clearly observe in Fig.~\ref{Fig-7}(\subref{EEexp}). However, the behavior of $\varepsilon_{\omega t}$ in Figs.~\ref{Fig-7}(\subref{EEHB},\subref{EER},\subref{EEE}) demonstrates that for large values of entropy, BH radiates permanently. The variation in the $\varepsilon_{\omega t}$ led BHs in high-energy configurations to emit radiation of low frequency with large values of entropy, which makes their identification very complicated. These phenomena are quite significant in the observational astronomy of BH evaporation for differentiating between the traditional general relativity and other gravity frameworks. 

Furthermore, from our analysis, it can be inferred that quantum gravitational corrections incorporated through the exponential corrected entropy suppressed strong thermal fluctuations and decreased the sparsity of Hawking radiation, effectively serving as a mechanism for information preservation via strong horizon correlation that slows down the evaporation of BHs.   

\section{Frequency Shift}\label{FRS} 
This section transitions our focus from thermodynamic to observational analysis through a study of the frequency shift experienced by test particles orbiting charged torus-like BHs. Moreover, we derive some basic expressions that are essential in this study to examine the frequency shift, for example, we need to derive geodesics for timelike and null particles by adopting the methodology given in Ref.~\cite{Martinez-Valera:2023guj} in our case (charged torus-like BH)\footnote{Here one can notice that the components construction in our case is significantly different from the spherically symmetric spapcetime given in Ref.~\cite{Martinez-Valera:2023guj}, for example in spherically symmetric spacetime we have $g_{\phi\phi}=r^{2}\sin^{2}\!\theta$ while in our case it is  $g_{\phi\phi}=r^{2}$ as dicussed in Refs.~\cite{Huang,Han,Ali:2023pyv,Feng:2021vey,Liang:2021elg}.}. We also want to highlight that all physical quantities are presented in natural (planks) units, making them dimensionless and ensuring that any comparison with astrophysical data remains qualitative. First, we define the line element, which we have already provided in Eq.~\eqref{1b} by utilizing $\Lambda=-3/l^2$, and the metric components with respect to this line element can be written as 
\begin{eqnarray}\nonumber
 g_{tt}&=&-\left(-\frac{2M}{\pi\ r}+\frac{4 Q^{2}}{\pi\ r^{2}}+\frac{r^{2}}{l^{2}}\right),\quad g_{rr}=\left(-\frac{2M}{\pi\ r}+\frac{4 Q^{2}}{\pi\ r^{2}}+\frac{r^{2}}{l^{2}}\right)^{-1}, \quad g_{\theta\theta}=r^{2},\\\label{MC}&& g_{\phi\phi} =  r^{2}\,.
\end{eqnarray}
\subsection{Massive Particles}
As we know, the geodesic equation for massive particles as described in Ref.~\cite{Martinez-Valera:2023guj} and its explicit form can be given as 
\begin{eqnarray}\label{GEMP}
 g_{\mathfrak{uv}}=-\mathbb{U}^{\mathfrak{u}}\mathbb{U}^{\mathfrak{v}}=-1\,,
\end{eqnarray}
where particle's four velocity is describe by $\mathbb{U}^{\mathfrak{u}}$. In order to obtain the conserved quantities, one can employ the killing vectors $\zeta^{\mathfrak{u}}=(1,~0,~0,~0)$ for timelike vector field  and $\Psi^{\mathfrak{u}}=(0,~0,~0,~1)$ for rotational vector field and by using this killing vectors we  obtain the conserved quantities given as 
\begin{eqnarray}\label{GEMP1}
 \mathbb{E}=\frac{\mathcal{E}}{m}=-\zeta_{\mathfrak{u}}\mathbb{U}^{\mathfrak{u}}=-g_{tt}\mathbb{U}^{\mathfrak{u}}, \quad \mathbb{L}=\frac{\mathcal{L}}{m}=\Psi_{\mathfrak{v}}\mathbb{U}^{\mathfrak{v}}=g_{\phi\phi}\mathbb{U}^{\mathfrak{v}}\,,
\end{eqnarray}
where $\mathbb{E}$ represents the total energy per unit mass, and $\mathbb{L}$ represents the angular momentum per unit mass of the test particle. If the particle is assumed to move within the equatorial plane, implying $\mathbb{U}^{\theta}=0$ and plugging these equations into Eq.~\eqref{GEMP}, which gives us the following relation
\begin{eqnarray}\label{GEMP2}
 -\left\{\frac{1}{2}g_{tt}g_{rr}(\mathbb{U}^{r})^{2}+\frac{g_{tt}}{2}+\frac{g_{tt}}{2g_{\phi\phi}}\mathbb{L}^{2}\right\}=\frac{\mathbb{E}^{2}}{2}\,.
\end{eqnarray}
It resembles the structure of the energy conservation law, in which the first term represents the moving particles' kinetic energy  within an effective potential that takes the following shape
\begin{eqnarray}\label{GEMP3}
\mathcal{V}_\mathrm{eff}=-\frac{1}{2}g_{tt}\left(1+\frac{\mathbb{L}^{2}}{g_{\phi\phi}}\right)\,.
\end{eqnarray}
One can notice that the relation given in Eq.~\eqref{GEMP2} only depends on the coordinate $r$, and it can be readjusted as follows
\begin{eqnarray}\label{GEMP4}
g_{rr}(\mathbb{U}^{r})^{2}=-1-\frac{\mathbb{L}^{2}}{g_{\phi\phi}}-\frac{\mathbb{E}^{2}}{g_{tt}}=\mathcal{V}_{r}(r)\,.
\end{eqnarray}
%spherical symmetry of
In light of the axisymmetric, toroidal spacetime and the characteristics of actual astrophysical systems, it is essential to examine equatorial circular motion. Putting $\theta=\text{constant}$\footnote{It is worthy to mention that we do not need to put $\theta=\frac{\pi}{2}$ which is common in the cases of spherically symmetric spacetime as described in Ref.~\cite{Martinez-Valera:2023guj}, the reason for this procedure is that our metric components are independent of the term $\sin\!\theta$. Therefore, we don't need to fix $\theta$, but one can fix it as described in Ref.~\cite{Ali:2023pyv}.} enables us to simplify a particular expression without any loss of generality, in our case. The motion of massive particles in circular orbits is obtained by following the conditions 
\begin{eqnarray}\label{GEMP5}
\mathcal{V}_{r}=-1-\frac{\mathbb{L}^{2}}{g_{\phi\phi}}-\frac{\mathbb{E}^{2}}{g_{tt}}=0\,,\quad \mathcal{V}_{r}'=-\frac{\mathbb{L}^{2}}{g_{\phi\phi}^{2}}g_{\phi\phi}'-\frac{\mathbb{E}^{2}}{g_{tt}^{2}}g_{tt}'=0\,,
\end{eqnarray}
where prime $'$ shows the derivative with respect to the radial coordinate $r$. The above-mentioned condition allows us to write $\mathbb{E}$ and $\mathbb{L}$ in the form of $g_{tt}$,~ $g_{\phi\phi}$, respectively, which takes the following shape  
\begin{eqnarray}\label{GEMP6}
\mathbb{E}^{2}=\frac{g^{2}_{tt}g_{\phi\phi}'}{g_{\phi\phi}g_{tt}'-g_{\phi\phi}' \ g_{tt}}\,,\quad \mathbb{L}^{2}=\frac{g_{tt}' \ g_{\phi\phi}^{2}}{g_{\phi\phi}' \ g_{tt}-g_{\phi\phi} \ g_{tt}'}\,.
\end{eqnarray}
By using Eq.~\eqref{GEMP6}, it  is easy to derive the energy and angular momentum of the particles of charged torus-like BHs, which turn out to be 
\begin{eqnarray}\label{ETBH}
 \mathbb{E}^{2}&=&-\frac{\left(l^2 \left(4 Q^2-2 M r\right)+\pi  r^4\right)^2}{\pi  l^4 r^2 \left(3 M r-8 Q^2\right)}\,,\\\label{LTBH}    \mathbb{L}^{2}&=&\frac{l^2 r^2 \left(M r-4 Q^2\right)+\pi  r^6}{l^2 \left(-2 M r+4 Q^2+\pi  r^4\right)+\pi  r^4}\,.
\end{eqnarray}

Thereby, it is straightforward to obtain the 4-velocity with respect to the temporal component $(t)$ and toroidal component $\phi$, which can  be written as 
\begin{eqnarray}\label{GEMP7}
\mathbb{U}^{t}_\mathrm{em}=-\frac{\mathbb{E}}{g_{tt}}= \sqrt{\frac{r^2\pi}{8 Q^2-3 M r}}\,,\quad \mathbb{U}^{\phi}_\mathrm{er}=\frac{\mathbb{L}}{g_{\phi\phi}}=\sqrt{\frac{l^2 \left(M r-4 Q^2\right)+\pi  r^4}{l^2 r^2 \left(8 Q^2-3 M r\right)}}\,.
\end{eqnarray}
Here, we mention that in the subscript of the 4-velocity, $\mathrm{er}$ is the radius of the emitter. In order to obtain the innermost stable circular orbit (ISCO), one can take the second-order derivative of Eq.~\eqref{GEMP5}, which is given as 
\begin{eqnarray}\label{GEMP8}
V''(r)=-\frac{2\mathbb{E}^{2}}{g_{tt}^{3}}(g'_{tt})^{2}+\frac{\mathbb{E}^{2}}{g_{tt}^{2}}g''_{tt}-\frac{2\mathbb{L}^{2}}{g_{\phi\phi}^{3}}(g''_{\phi\phi})^{2}+\frac{\mathbb{L}^{2}}{g_{\phi\phi}^{2}}g''_{\phi\phi}\leq 0\,.
\end{eqnarray}
The equality here corresponds to the ISCO that defined the inner boundary of the accretion disk as described in Ref.~\cite{Martinez-Valera:2023guj}. By using Eq.~\eqref{GEMP6}, we can explicitly present Eq.~\eqref{GEMP8} in the following form 
\begin{eqnarray}\label{GEMP9}
V''(r)=\frac{1}{g_{\phi\phi}' \ g_{tt}-g_{\phi\phi} \ g_{tt}'}\left[g'_{\phi\phi}\left\{\frac{2(g'_{tt})^{2}}{g_{tt}}-g''_{tt}\right\}-g'_{tt}\left\{\frac{2(g'_{\phi\phi})^{2}}{g_{\phi\phi}}-g''_{\phi\phi}\right\}\right]\,.
\end{eqnarray}

By using Eq.~\eqref{GEMP9}, one can compute the $r_\mathrm{ISCO}$ for charged torus-like BH, which is given as 
\begin{eqnarray}\label{RISCOo}
  -\frac{2 \left(l^2 \left(6 M^2 r^2-36 M Q^2 r+64 Q^4\right)+3 \pi  r^4 \left(5 M r-16 Q^2\right)\right)}{r^2 \left(3 M r-8 Q^2\right) \left(l^2 \left(4 Q^2-2 M r\right)+\pi  r^4\right)}= 0\,. 
\end{eqnarray}

It is difficult to obtain $r_\mathrm{ISCO}$ analytically from Eq.~\eqref{RISCOo} due to the presence of the fifth-order polynomial. Therefore, we numerically obtain the values of $r_\mathrm{ISCO}$ for different values of $\Lambda$, which we presented in Table \ref{tab:riscocomparison}, and compare our analysis to other classical solutions such as Schwarzschild and RN-AdS BHs.
\subsection{Null Particles}
The equation of motion for massless particles (namely photons) can be presented as 
\begin{eqnarray}\label{NG}
g^{\mathfrak{u}\mathfrak{v}}\mathbb{K}^\mathfrak{u}\mathbb{K}^\mathfrak{v}=0\,,
\end{eqnarray}
where the photon's 4-wave vector is presented  by $\mathbb{K}^\mathfrak{u}=\mathbb{K}^{t},\mathbb{K}^{r},\mathbb{K}^{\theta},\mathbb{K}^{\phi}$. The motion of these massless particles propagating outside the horizon is described by Eq.~\eqref{1b}, and it depends on the explicit form of the spacetime component mentioned in Eq.~\eqref{MC}. Since photons are assumed to move within the equatorial plane, the component of the 4-wave vector $\mathbb{K}^{\theta}$ becomes zero, which leads us to the following expression
\begin{eqnarray}\label{NG1}
g_{tt}(\mathbb{K}^{t})^{2}+g_{rr}(\mathbb{K}^{r})^{2}+g_{\phi\phi}(\mathbb{K}^{\phi})^{2}=0\,.
\end{eqnarray}
By adopting a similar methodology to that adopted  for spherically symmetric spacetime in Ref.~\cite{Martinez-Valera:2023guj} to our case (charged BH in toroidal spacetime), the energy and angular momentum of the photon particles can be written as 
\begin{eqnarray}\label{NG2}
\mathbb{E}_{\gamma}&=&-\zeta_\mathfrak{u}\mathbb{K}^\mathfrak{u}=g_{tt}\mathbb{K}^{t}\,,\\\label{NG3} \mathbb{L}_{\gamma}&=&-\Psi_\mathfrak{u}\mathbb{K}^\mathfrak{u}=g_{\phi\phi}\mathbb{K}^{\phi}\,. 
\end{eqnarray}
By using Eqs.~\eqref{NG2} and \eqref{NG3} in Eq.~\eqref{NG1}, which yields
\begin{eqnarray}\label{NG4}
g_{rr}(\mathbb{K}^{r})^{2}+\frac{\mathbb{E}_{\gamma}}{g_{tt}}+\frac{\mathbb{L}_{\gamma}}{g_{\phi\phi}}=0\,.
\end{eqnarray}

It is important to note that at points where the radial component of the 4-wavevector $\mathbb{K}^{r}$ becomes zero, particularly at the diametrically opposite location along a line perpendicular to the line of sight. Therefore, Eq.~\eqref{NG4} is reduced to the given form 
\begin{eqnarray}\label{NG5}
\frac{\mathbb{E}_{\gamma}}{g_{tt}}+\frac{\mathbb{L}_{\gamma}}{g_{\phi\phi}}=0\,.
\end{eqnarray}
It is straightforward to obtain the parameter for the bending of light by employing Eq.~\eqref{NG5}, which gives
\begin{eqnarray}\label{NG6}
b_{\gamma}=\frac{\mathbb{L}_{\gamma}}{\mathbb{E}_{\gamma}}\,,
\end{eqnarray}
which can be further simplified in terms of spacetime components
\begin{eqnarray}\label{NG7}
b_{\gamma}=\pm\sqrt{\frac{g_{\phi\phi}}{g_{tt}}}=\pm \sqrt{\pi } \sqrt{\frac{l^2 r^4}{l^2 \left(4 Q^2-2 M r\right)+\pi  r^4}}\,.
\end{eqnarray}

This relation provides information about the deflection of light sources located on either side of the BH along the midline, as indicated by the sign $\pm$.

Now, we examine photons frequency located at the $\mathfrak{x}^{\mathfrak{u}}_\mathrm{p}=(\mathfrak{x}^{\mathfrak{t}},~\mathfrak{x}^{r},~\mathfrak{x}^{\theta},~\mathfrak{x}^{\phi})$ given as 
\begin{eqnarray}\label{FS}
\omega_\mathrm{p}=-\mathbb{K}_{\mathfrak{u}}\mathbb{U}^{\mathfrak{u}}\bigg|_\mathrm{p}\,,
\end{eqnarray}
where photon's point of emission $\mathfrak{x}^{\mathfrak{u}}_\mathrm{em}$ or detection $\mathfrak{x}^{\mathfrak{u}}_\mathrm{dc}$ is presented here by $\mathrm{p}$. By adopting the same process as described in Refs.~\cite{Herrera-Aguilar:2015kea,Banerjee:2022him,Momennia:2023lau,Martinez-Valera:2023guj}, it is easy to define the general expression of  frequency shift for the axisymmetric toroidal spacetime, which can be written as 
\begin{eqnarray}\label{FS1}
1+z_\mathrm{BH}=\frac{\omega_\mathrm{em}}{\omega_\mathrm{dc}}=\frac{\left(\mathbb{E}_{\gamma}\mathbb{U}^{t}-\mathbb{L}_{\gamma}\mathbb{U}^{\phi}-g_{rr}\mathbb{U}^{r}\mathbb{K}^{r}-g_{\theta\theta}\mathbb{U}^{\theta}\mathbb{K}^{\theta}\right)\bigg|_\mathrm{em}}{\left(\mathbb{E}_{\gamma}\mathbb{U}^{t}-\mathbb{L}_{\gamma}\mathbb{U}^{\phi}-g_{rr}\mathbb{U}^{r}\mathbb{K}^{r}-g_{\theta\theta}\mathbb{U}^{\theta}\mathbb{K}^{\theta}\right)\bigg|_\mathrm{dc}}\,.
\end{eqnarray}
In addition, we set $\mathbb{U}^{\theta}=0$, owing to the fact that the motion of massive test particles is limited to equatorial plane and by using the following condition $\mathbb{U}^{r}=0$ for the circular orbit, the expression given in Eq.~\eqref{FS1} takes the following form 
\begin{eqnarray}\label{FS2}
1+z_\mathrm{BH}=\frac{\left(\mathbb{E}_{\gamma}\mathbb{U}^{t}-\mathbb{L}_{\gamma}\mathbb{U}^{\phi}\right)\bigg|_\mathrm{em}}{\left(\mathbb{E}_{\gamma}\mathbb{U}^{t}-\mathbb{L}_{\gamma}\mathbb{U}^{\phi}\right)\bigg|_\mathrm{dt}}\,,
\end{eqnarray}
We can further simplify Eq.~\eqref{FS2} by assuming 4-velocity $\mathbb{U}^\mathfrak{u}_\mathrm{dc}=(1,~0,~0,~0)$ as the distance between the observer and the BH is very large. Thereby, the above-mentioned equation can be written as  
\begin{eqnarray}\label{FS3}
1+z_\mathrm{BH}=\left(\mathbb{U}^{t}-b_{\gamma\pm}\mathbb{U}^{\phi}\right)\bigg|_\mathrm{em}\,.
\end{eqnarray}
If we examine Eq.~\eqref{FS3}, the $\pm$ signs reflect the direction of mass; for instance, a positive sign indicates clockwise rotation, while a negative sign indicates counterclockwise motion. Moreover, it is irrelevant whether the rotation of the particle is clockwise or counterclockwise in the cases of spherically symmetric spacetime and axisymmetric toroidal spacetime. Now, the general expression as given in Refs.~\cite{Herrera-Aguilar:2015kea,Banerjee:2022him,Momennia:2023lau,Martinez-Valera:2023guj} can be obtained by utilizing Eqs.~\eqref{GEMP7},~\eqref{NG5}, and \eqref{FS3}, which can be written as 
\begin{eqnarray}\nonumber
1+z_\mathrm{BH_{_{_{_{1,2}}}}}&=&\frac{\sqrt{g_{\phi\phi}'} \ \pm \sqrt{\frac{g_{\phi\phi}}{g_{tt}}g_{tt}'} }{g_{\phi\phi}g_{tt}'-g_{\phi\phi}' \ g_{tt}} =\sqrt{\pi } \Bigg(\sqrt{\frac{r^2_\mathrm{e}}{8 Q^2-3 M r_\mathrm{e}}}\pm\sqrt{\frac{l^2 \left(M r_\mathrm{e}-4 Q^2\right)+\pi  r^4_\mathrm{e}}{l^2 r^2_\mathrm{e} \left(8 Q^2-3 M r_\mathrm{e}\right)}} \\\label{FS4}&\times&\sqrt{\frac{l^2 r^4_\mathrm{e}}{l^2 \left(4 Q^2-2 M r_\mathrm{e}\right)+\pi  r^4_\mathrm{e}}}\Bigg)\,.
\end{eqnarray}
where the shift of the photon's frequency owing to circular orbiting sources in a spherically symmetric spacetime is obtained at the horizon radius $r_\mathrm{e}$. The $\pm$ sign indicates the association between  $z_\mathrm{BH_{1}}$ and $z_\mathrm{BH_{2}}$, which corresponds to the redshift and blueshift at the central line and along both directions from the line of sight. The above equation reflects the total shift of frequency whereas the first term corresponds to the gravitation shift of frequency and the second term is linked with kinetic frequency shift as following expression $z_\mathrm{BH_{_{_{_{1,2}}}}}=z_\mathrm{g}+z_{\mathrm{ke}_{\pm}}$ given in Ref.~\cite{Martinez-Valera:2023guj}. From Eq.~\eqref{FS4}, we can identify the gravitational frequency shift and kinetic energy frequency  shift, which are given as 
\begin{eqnarray}\label{FSG}
1+z_\mathrm{g}&=&\frac{\sqrt{r_\mathrm{e} \ \pi}}{\sqrt{\frac{8 Q^2}{r_\mathrm{e}}-3 M}}\,,\\\label{FSKE} z_\mathrm{ke_{\pm}}&=&\pm \sqrt{\pi } \sqrt{\frac{l^2 r^4_\mathrm{e}}{l^2 \left(4 Q^2-2 M r_\mathrm{e}\right)+\pi  r^4_\mathrm{e}}} \sqrt{\frac{l^2 \left(M r_\mathrm{e}-4 Q^2\right)+\pi  r^4_\mathrm{e}}{l^2 r^2_\mathrm{e} \left(8 Q^2-3 M r_\mathrm{e}\right)}}\,. 
\end{eqnarray}

\begin{figure} [t]
\epsfig{file=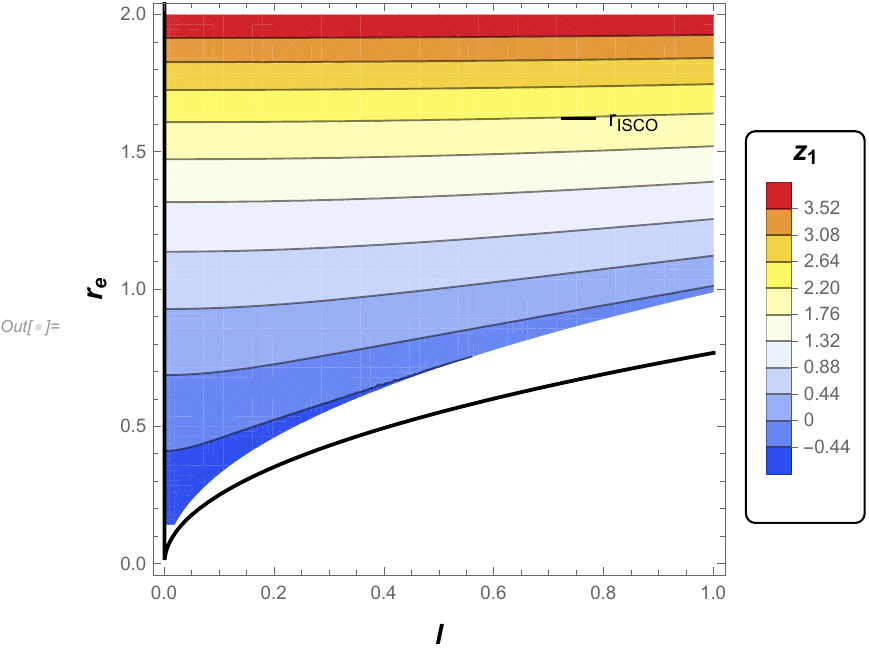,width=.45\linewidth} 
\epsfig{file=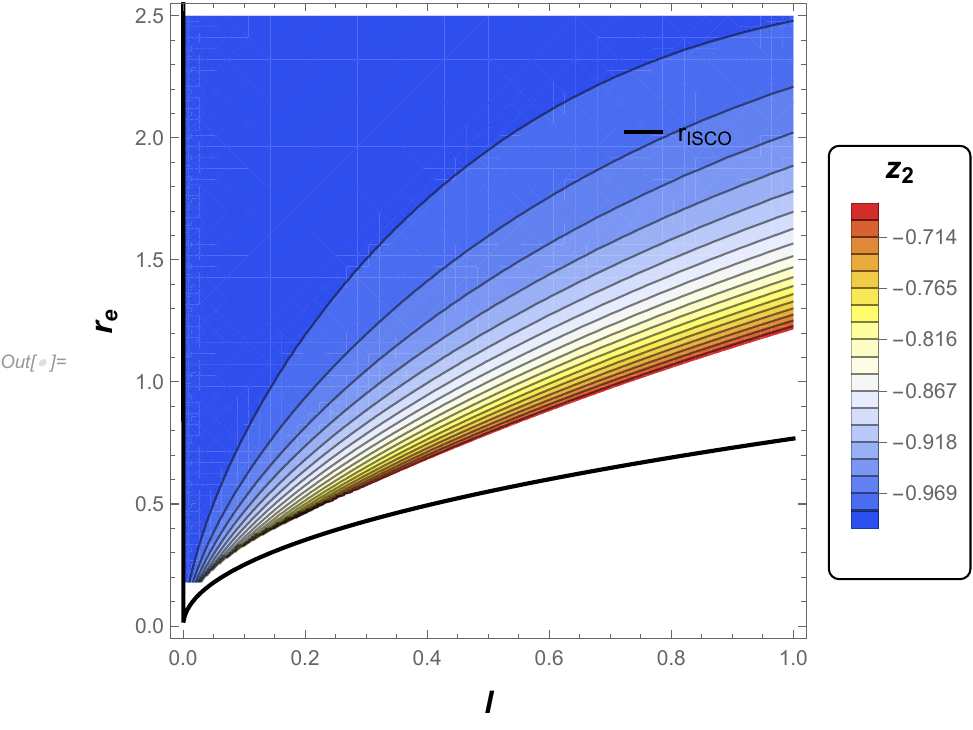,width=.45\linewidth} \caption{\raggedright The redshift $z_{1}$ (left panel) and the blueshift (right panel) in $r_\mathrm{e}-l$ plane by setting $M=1$ and $Q=1$. In addition, the black curve presents $r_\mathrm{e}=r_\mathrm{ISCO}$, which we numerically compute. }\label{Fig-8}
\end{figure}

\begin{figure} [ht]
\epsfig{file=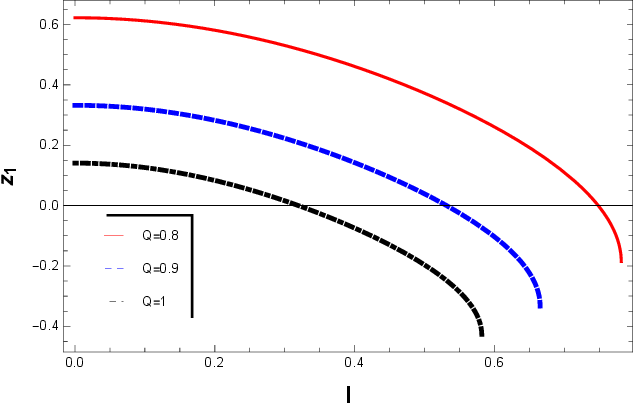,width=.45\linewidth} 
\epsfig{file=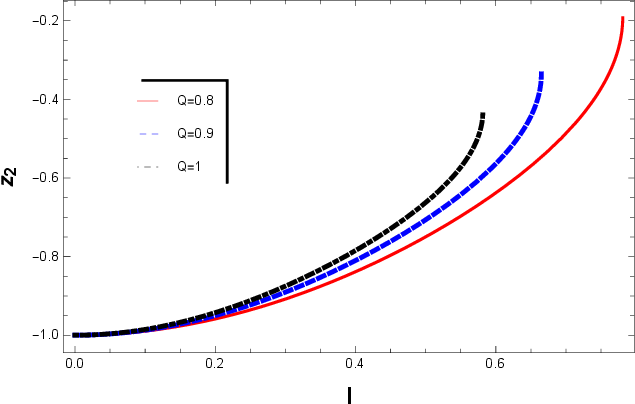,width=.45\linewidth}
\epsfig{file=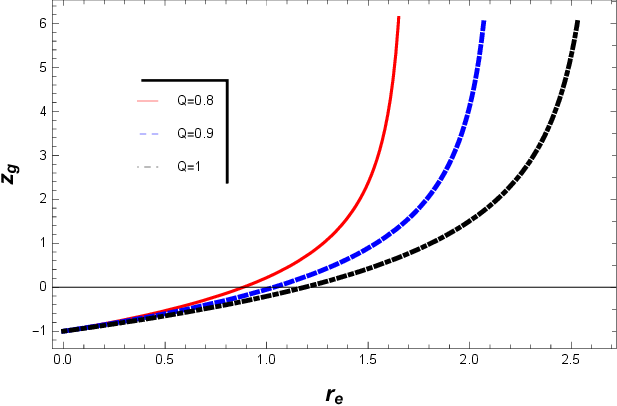,width=.45\linewidth} \caption{\raggedright Graphs of the redshift (upper left panel) and blueshift (upper right panel) as the function of $l$ by substituting $M=1,~r_\mathrm{e}=r_\mathrm{ISCO}$ and for different trajectories we varied the electric charge $Q$ such as $Q=0.8$ (red curve), $Q=0.9$ (blue dotted curve) and $Q=1$ (black dotted-dashed curve). We also mention that since the gravitational shift (lower panel) is free from the AdS radii $l$, we demonstrate $z_\mathrm{g}$ in terms of $r_\mathrm{e}$.}\label{Fig-9}
\end{figure}

We graphically present the behavior of redshift (left panel) and blueshift (right panel) by putting $M=1,~Q=1$ in the $r_\mathrm{e}-l$  plane. Additionally, it's worth noting that we select $r_\mathrm{e}-l$ to cover the range $l\in[0,1]$ and maintain consistency with our study. Moreover, the black curve in Fig.~\ref{Fig-8} is $r_\mathrm{ISCO}$, which we obtain by numerically solving Eq.~\eqref{RISCOo}, and it also satisfies the condition of orbit stability $r_\mathrm{ISCO}\leq r$. We observe that in the case of a charged torus-like BH $r_\mathrm{ph}$, which is obtained by using the condition provided in Ref.~\cite{Martinez-Valera:2023guj}, violates $r_\mathrm{ph}\nleq r_\mathrm{ISCO}$. The reason for this violation is that the metric function for charged torus-like BH is different from the Schwarzschild case for example in our case we compute $r_\mathrm{ph}=\frac{8Q^{2}}{3M}$ (which also agrees with the one given in Ref.~\cite{Ali:2023pyv}), it indicates that $r_\mathrm{ph}$ is inversely in relation with the mass and directly proportional with $Q$. In the case of the Schwarzschild BHs, we have $r_\mathrm{ph}=3M$. Thereby, to satisfy the condition $r_\mathrm{ph}<r_\mathrm{ISCO}$, if we take $M\geq1$  and $Q\leq1$. We observe in Fig.~\ref{Fig-8} that there are opposite impacts of $r_\mathrm{e}$ and $l$ on redshift and blueshift frequencies, for example, as $r_\mathrm{e}$ and $l$ increase, the frequency of blueshift (left panel) and redshift (right panel) also increases (with negative behavior) and decreases (with positive behavior), respectively. Consequently, we assume that these parameters can balance the frequency shift at the points marked by the grey curves, which represent constant contour levels $z_{1,2}$.

\begin{table}[t]
\caption{\label{tab:riscocomparison}\raggedright Summary of the numerical values of $r_\mathrm{ISCO}$ and also its comparison with the RN-AdS and Schwarzschild BHs. One can also verify the condition violation of $r_\mathrm{ph}< r_\mathrm{ISCO}$ for $Q > 1$, while we compared it in the case of spherically symmetric spacetime, such as RN-AdS and Schwarzschild BHs. We can clearly observe that the violation which we highlighted is also observed in some cases of RN-AdS BHs. We also mention here that CTL BHs present the abbreviation of charge torus-like BH.}
\begin{ruledtabular}
\small
\begin{tabular}{cccc}
\textbf{Values}    & \textbf{CTL BH} & \textbf{RN-AdS BH} & \textbf{Schwarzschilld BH}\\
\hline 
 & $r_\mathrm{ph}$~~~~~~$r_\mathrm{ISCO}$ & $r_\mathrm{ph}$~~~~~~$r_\mathrm{ISCO}$ & ~~~~~$r_\mathrm{ph}$~~~~~$r_\mathrm{ISCO}$ \\
\hline
$Q=0.1,~l=0.1$            & 0.026667~~~0.178122                &2.99332~~~3.74209         &3~~~~~~6                      
\\ $Q=0.1,~l=0.5$          & 0.026667~~~0.535088              & 2.99332~~~3.74464     &3~~~~~~6                         \\
$Q=0.1,~l=1$          & 0.026667~~~0.853481              & 2.99332~~~3.75253 &3~~~~~~6                   \\ \\
$Q=0.5,~l=0.1$            & 0.666667~~~0.799668                &2.82288~~~3.53813            &3~~~~~~6             \\
$Q=0.5,~l=0.5$          & 0.666667~~~0.791409              & 2.82288~~~3.54078                  &3~~~~~~6            \\
$Q=0.5,~l=1$          & 0.666667~~~0.76026              & 2.82288~~~3.54894            &3~~~~~~6        \\
              \\
$Q=1,~l=0.1$            & 2.66667~~~3.19998                & 1.5~~~~0.225115            &3~~~~~~6             \\
$Q=1,~l=0.5$          & 2.66667~~~3.19948              & 1.5~~~~2.59649                  &3~~~~~~6            \\
$Q=1,~l=1$          & 2.66667~~~3.19792              & 1.5~~~~2.60651            &3~~~~~~6        \\
              \\
$Q=1.1,~l=0.1$            & 3.22667~~~0.264025                & 1.24206~~~0.238076            &3~~~~~~6             \\
$Q=1.1,~l=0.5$          & 3.22667~~~0.580468              & 1.24206~~~0.488289                  &3~~~~~~6            \\
$Q=1.1,~l=1$          & 3.22667~~~0.810754              & 1.24206~~~0.638129            &3~~~~~~6        \\
\end{tabular}
\end{ruledtabular}
\end{table}

We present the behavior of redshift $z_{1}$ (upper left panel), blueshift $z_{2}$ (upper right panel) and gravitation shift $z_\mathrm{g}$ (lower panel) by setting  $M=1,~r_\mathrm{e}=r_\mathrm{ISCO}$ and compute various curves for different values of $Q$ such as for $Q=0.8$ (red curve), $Q=1$ (blue dotted curve) and $Q=1.3$ (black dot-dashed curve). It is evident from Fig.~\ref{Fig-9} that increasing the parameter $Q$ leads to a transition from higher to lower values of the redshift $z_{1}$ (upper left panel), whereas in the case of blueshift $z_{2}$ (upper right panel), it increases for different values of charge $Q$. It also indicates that, as the parameter $Q$ increases, the frequency-shift curves grow more rapidly with $l$. Notably, the gravitational shift $z_\mathrm{g}$ is initially negative but turns positive and increase with $Q$ as the AdS radii $l$ grows as shown in Fig.~\ref{Fig-9} (lower panel), while on the other hand $z_{2}$ (upper right panel) becomes increasingly negative with rising $Q$, owing to the impact of $z_{\mathrm{kin}^{-}}$ component. Furthermore, from the relation of gravitation shift, we observe that it is independent of the term $l$, so we graphically present the behavior of $z_\mathrm{g}$ in the form of $r_\mathrm{e}$, which indicates that for large values of $r_\mathrm{e}$, gravitational shift obtains maximum values. Thereby, the curves in Fig.~\ref{Fig-9} represent the highest frequency shifts, computed for an emitter revolving circularly around the BH at $r_\mathrm{e}=r_\mathrm{ISCO}$, and also as the orbital radius grows infinitely, then the frequency shift vanishes to zero. One can plugged $R=1+z_{1}$ and $B=1+z_{2}$  to verify 
\begin{eqnarray}\label{FSG1}
RB&=&\frac{\pi  l^2 r^2_\mathrm{e}}{l^2 \left(4 Q^2-2 M r_\mathrm{e}\right)+\pi  r^4_\mathrm{e}}~. 
\end{eqnarray}
This expression yields an expression for the mass of the non-singular BHs depending on the values of $z_{1}$ and $z_{2}$, which takes the following shape
\begin{eqnarray}\nonumber
M&=&\frac{4 l^2 Q^2+\pi  r^4_\mathrm{e}}{2 l^2 r_\mathrm{e}}-\frac{\pi  r_\mathrm{e}}{2 \text{RB}}\,. 
\end{eqnarray}
The highest values of redshift $z_{1}$ and blueshift $z_{2}$ for revolving particles in ISCO are obtained at $r_\mathrm{ISCO}$. One can rely on observational data of redshift $z_{1}$ and blueshift $z_{2}$ derived from $\mathrm{H_{2}O}$ megamaser present in the accretion disk surrounding supermassive BHs with active galactic nuclei cores; for example in the case of NGC 4258, measurement from its central megamaser system demonstrate that the redshift range lies in $z_{1}\in (4\times 10^{-3},~6\times 10^{-3})$, blueshift range lies in $z_{2}\in (-1.7\times 10^{-3},~-1.0\times 10^{-3})$ and the emitter radius is located in the range of $r_\mathrm{e}\in(0.04pc,~0.5pc)$ from the center of the disk \cite{Herrnstein:2005xc}. The data computed from megamaser glaxies NGC 1194, NGC 2273, NGC 2960, NGC 6264, NGC 6323, and UGC 3789 indicates the redshift range lies in  $z_{1}\in (1.1\times 10^{-3},~3.3\times 10^{-3})$, blueshift range is $z_{2}\in (-2.6\times 10^{-3},~-0.9\times 10^{-3})$ and the emitter radius lies in $r_\mathrm{e}\in(0.028pc,~1.33pc)$ \cite{Villaraos:2022euo,Reid:2008nm,Kuo:2010uy}. The gravitational redshift in megamaser systems has been shown to range from $z_\mathrm{g}\in(1.6\times 10^{-7},~3.1\times 10^{-5})$, governed by the mass of the central supermassive BH and the distance to the emitting region as discussed in Refs.~\cite{Nucamendi:2020tov,Villalobos-Ramirez:2022xic,Villaraos:2022euo,Gonzalez-Juarez:2022pya}. Thereby, one can employ the generic relativistic method to an astrophysical system to compute the gravitational redshift $z_\mathrm{g}$.     

\section{Conclusions}\label{CNC}
In this work, we have comprehensively discussed the thermodynamic and observational analysis for a charged torus-like BH. This paper is structured in two parts; the initial part focuses on the impact of different entropy frameworks, such as the Hawking-Bekenstein entropy, exponential corrected, and R\'{e}nyi entropies, on the thermodynamics of a charge torus-like BH, while the latter part focuses on the observational aspects determined purely from the spacetime geometry. In the thermodynamic analysis, we have employed different entropy frameworks, such as the Hawking-Bekenstein, exponential corrected, and R\'{e}nyi entropies, to analytically compute various thermodynamic quantities such as mass $M$, temperature $T$, and thermodynamic volume $V$.  By utilizing these quantities, we computed and graphically presented the heat capacity, $P-V$ diagram, isotherm compressibility, and free energies (HFE and GFE). 
In this analysis, it is notably found that there is only one ZP of heat capacity in the case of the Hawking-Bekenstein and R\'{e}nyi, while on the other hand, the exponential corrected entropy demonstrates two distinct ZPs of heat capacity. This suggests that exponentially corrected entropy provides a more intricate thermal phase structure and more complex stability behavior.  To validate the physical significance of these identified ZPs, we employed Weinhold and Ruppeiner geometries, observing that the Ricci scalar obtained from the Ruppeiner geometry diverges at the same point where heat capacity becomes zero, thereby providing geometric validation of the phase transition. Compared to the Hawking-Bekenstein and R\'{e}nyi entropies, where free energies (HFE and GFE) demonstrate the fluctuating behavior indicating instability, the exponential corrected entropy approach leads to positive and stable behavior in both HFE and GFE, highlighting global thermodynamic stability.  Similarly, the exponential corrected entropy model shows a non-monotonic pressure trend in the $P-V$ diagram, indicating a possible phase transition absent in other entropy models, and this is also confirmed by the behavior of isothermal compressibility. Additionally, we presented the impact of entropy corrections on the thermodynamics of charged torus-like BH by comparing it with the spherically symmetric spacetimes such as Schwarzschild and RN-AdS BHs in Table~\ref{tab:comparison}. 

Furthermore, analysis through the GTPE framework clarifies that exponentially corrected entropy exhibits significant thermodynamic deviations, highlighting its strong sensitivity to microstructures of the BHs as summarized in Table~\ref{tab:comparison}. Whereas, the Hawking-Bekenstein and R\'{e}nyi entropies align well with GPTE, signalling the same universality class; however, the exponential corrected entropy behaves differently, reflecting a non-perturbative modification. This highlights that, as compared to other entropy models, the exponentially corrected entropy uniquely captures subtle microstructural aspects linked to the toroidal horizon topology. Our analysis reveals that the choice of entropy plays a pivotal role in determining the thermodynamics of BHs in toroidal spacetime, highlighting the ability of most generalized entropies to investigate quantum and topological phenomena beyond the traditional area law. Thereby, this study offers a robust and physically motivated groundwork for examining non-perturbative and non-extensive entropy corrections in BH thermodynamics, and also suggests various possible directions for future work utilizing the generalized entropy (with six, five, or four parameters) framework as given in Refs.~\cite{Nojiri:2022aof,Nojiri:2022dkr,Nojiri:2022sfd,Elizalde:2025iku,Odintsov:2023vpj,Nojiri:2024zdu,Odintsov:2022qnn}. For example, a worthwhile future direction is to employ the generalized six parameters entropy via numerical inversion methods to further examine the universality among topological BHs, which could provide more valuable insights.

We further studied the sparsity of Hawking radiation and the energy emission rate by utilizing the conjugate temperature, which corresponds to the respective entropy models used in this analysis. It is observed that for small ranges of entropy, the parameter of sparsity $\eta$ increased, but for large values of entropy, it began to decrease. After some time, it became constant for the exponential corrected entropy, but in the case of the Hawking-Bekenstein and R\'{e}nyi entropy, it converges to zero. As we mentioned earlier, our aim in this study is to investigate the theoretical (thermodynamics) and observational characteristics of the charged torus-like BH. We analyzed the motion of massive and massless test particles via geodesics to examine the observational aspects of a charged torus-like BH.  We analytically determined the redshift, blueshift, and gravitational frequency shift of emitted photons originating from the orbital motion. It is notably observed that, for specific parameter choices such as $Q> 1$ and $M\leq 1$, the condition $r_\mathrm{ph}<r_\mathrm{ISCO}$ may not hold for charged torus-like BHs due to their spacetime geometry. We observed that this inequality might hold under the following conditions, such as $M>1$ and $Q\leq1$. Additionally, our analysis confirms that the calculated redshift $z_{1}$ and blueshift $z_{2}$ agree with the observational constraints for the megamaser disk systems NGC 4258 and UGC 3789. The observed consistency suggests that toroidal BH models may be suitable for studying specific astrophysical BHs, notably when examined through the lens of frequency-shift phenomena. Our analysis highlights the significant role of entropy correction in BH thermodynamics and identifies observational features arising from toroidal geometries. 

We want to extend this analysis in the future by exploring rotating toroidal BHs and higher-dimensional geometries to study the role of angular momentum and the impact of extra dimensions on thermodynamic and observational features. In addition, one can employ different quantum-corrected entropy models, such as the Barrow and Sharma-Mittal entropies, to investigate the thermodynamics of BHs, which may offer valuable insights into their stability and phase transitions. From an observational perspective, investigating lensing, shadow, and quasinormal modes in torus-like geometries and comparing them with the recent LIGO-Virgo \cite{LIGOScientific:2016aoc,LIGOScientific:2016sjg} and EHT data \cite{EventHorizonTelescope:2019ths,EventHorizonTelescope:2022xqj} may enhance the precision of BH parameter constraints. 
\begin{acknowledgments}
The work of K.B. was supported by the JSPS KAKENHI Grant Numbers 24KF0100, 25KF0176, and Competitive Research Funds for Fukushima University Faculty (25RK011).
\end{acknowledgments}

\end{document}